\documentclass[twocolumn]{aastex62}

\usepackage{amssymb}
\usepackage{amsmath}
\usepackage{mathtools}
\usepackage{graphicx}
\usepackage{color}
\usepackage{nicefrac}
\usepackage{bm, xspace}

\def\Jc{J_{2, {\rm crit}}}
\def\td{t_{\rm diff}}
\def\te{t_{\rm e-fold}}
\def\Md{M_{\rm d}}
\def\rc{r_{\rm crit}}

\renewcommand{\c}[2]{\ensuremath{#1_{\text{#2}}}} 	
\newcommand{\ia}{\c{i}{a}\xspace}
\newcommand{\ib}{\c{i}{b}\xspace}
\newcommand{\ie}{\c{i}{e}\xspace}
\newcommand{\unitvectorslope}[1]{\ensuremath{%
\frac{\c{\hat{#1}}{z}}%
     {\sqrt{1-\c{\hat{#1}}{z}^2}}}}

\begin{document}

\title{Giant Planet Influence on the Collective Gravity of a Primordial Scattered Disk}

\author{Alexander Zderic}
\affiliation{JILA and Department of Astrophysical and Planetary Sciences, CU Boulder, Boulder, CO 80309, USA}
\email{alexander.zderic@colorado.edu}

\author{Ann-Marie Madigan}
\affiliation{JILA and Department of Astrophysical and Planetary Sciences, CU Boulder, Boulder, CO 80309, USA}

\shorttitle{Giant planets and the Inclination Instability}
\shortauthors{Zderic \& Madigan}

\begin{abstract}

Axisymmetric disks of high eccentricity, low mass bodies on near-Keplerian orbits are unstable to an out-of-plane buckling. This ``inclination instability'' exponentially grows the orbital inclinations, raises perihelion distances and clusters in argument of perihelion. Here we examine the instability in a massive primordial scattered disk including the orbit-averaged gravitational influence of the giant planets. 
We show that differential apsidal precession induced by the giant planets will suppress the inclination instability unless the primordial mass is $\gtrsim 20$ Earth masses. We also show that the instability should produce a  ``perihelion gap'' at semi-major axes of hundreds of AU, as the orbits of the remnant population are more likely to have extremely large perihelion distances ($\mathcal{O}(100~\rm{AU})$) than intermediate values.

\end{abstract}


\section{Introduction}
\label{sec:intro}

Structures formed by the collective gravity of numerous low-mass bodies are well-studied on many astrophysical scales, for example, stellar bar formation in galaxies \citep{Sellwood1993} and apsidally-aligned disks of stars orbiting supermassive black holes \citep{Kazandjian2013, Madigan2018}. The driver of these dynamics are long-term (secular) gravitational torques between orbits.

The corresponding structures in planetary systems are relatively under-explored. This may be due to the presence of massive perturbers (planets) that are assumed to dominate the dynamics. While this is often the case on small scales close to the host star, there may be significant regions of phase space in which the influence of massive planets is small and the orbital period of bodies is short enough for collective gravitational torques to be important. 

In \citet{Madigan2016} we presented the discovery of a gravo-dynamical instability driven by the collective gravity of low mass, high eccentricity bodies in a near-Keplerian disk. This ``inclination instability'' exponentially grows the orbital inclination of bodies while decreasing their orbital eccentricities and clustering their arguments of perihelion ($\omega$). 

In \citet{Madigan2018} we explained the mechanism behind the instability: secular torques acting between the high eccentricity orbits. We also showed how the instability timescale scaled as a function of disk parameters. One important result is that the growth timescale is sensitive to the number of bodies used in $N$-body simulations. A low number of particles suppresses the instability due to two-body scattering and incomplete angular phase coverage of orbits in the disk. 

We showed that the amount of mass needed for the collective gravity of extreme trans-Neptunian objects (eTNOs) to be the dominant dynamical driver in the outer Solar System ($\sim100-1000$ AU) was about half an Earth mass. However, to observe significant clustering in $\omega$ within the age of the Solar System we required a mass closer to a few Earth masses. We note that this is very similar to the predicted mass of Planet 9 \citep{Batygin2016,Batygin2019}. It is perhaps no coincidence that the mass requirements are the same as dynamics are driven by gravitational torques in both \citep{Batygin2017}, but a disk of individually low mass bodies with high perihelion and inclinations will be harder to observe than a single massive body at the same distance.   
In \citet{Fleisig2020} we moved from simulations of a single mass population to a mass spectrum.
In this paper we add two more additional complexities to the system: a more realistic orbital configuration and the gravitational influences of the giant planets.
Our goal is to determine the parameters under which the presence of giant planets completely suppresses the inclination instability in a orbital configuration modeled on a primordial scattered disk \citep{Luu1997,Duncan1997}.  
We note that this was first addressed by \citet{Fan2017}, who found that the inclination instability did not occur in their simulations of the Nice Model containing $30$ Earth masses of self-gravitating planetesimals. These simulations, however, lacked a sufficient number of particles between 100 - 1000 AU ($N < 16$) for the inclination instability to occur. 
We find that differential apsidal precession induced by the giant planets can suppress the inclination instability in the scattered disk. However, if the mass of the primordial scattered disk is large enough ($\gtrsim 20$ Earth masses) then the instability will occur.

In Section~\ref{sec:sims}, we describe our $N$-body simulations including how we emulate the influence of the giant planets with a quadrupole ($J_2$) potential. In Section~\ref{sec:suppress}, we discuss how the instability is changed by the $J_2$ potential and show results for a primordial scattered disk configuration.
We also discuss the generation of a ``perihelion gap'' at hundreds of AU.
In Section~\ref{sec:scaling}, we scale our results to the solar system, obtaining an estimate for the required primordial mass of the scattered disk for the inclination instability to have occurred within it.  Finally, in Section~\ref{sec:sum}, we summarize our results and discuss the implications of our work.

\section{Numerical Methods}
\label{sec:sims}

\subsection{$N$-body Simulations}
To study the collective gravitational effects of minor bodies in the outer Solar System we run simulations using \texttt{REBOUND}, an open-source $N$-body integration framework available in C with a Python wrapper. \texttt{REBOUND} offers a few different integration methods and gravity algorithms \citep{Rein2012}. For this work, we use the direct gravity algorithm ($N^2$ scaling) and the IAS15 adaptive time-step integrator. 
We also use the additional package \texttt{REBOUNDx}  which provides a framework for adding additional physics (e.g. general relativity, radiation forces, user-defined forces) \citep{Tamayo2019}. 

\subsection{JSUN as $J_{\rm 2,Sun}$}
The most straight-forward way to incorporate the giant planets would be to simulate them directly as $N$-bodies. However, this is much harder to do than it might seem. Out of computational necessity we simulate implausibly \replaced{large (small) disk masses (number of particles)}{ large particle masses} and scale our results to realistic values. If we wanted to simulate the correct \replaced{{\it mass ratio}}{mass ratio} between giant planets and the disk, the mass ratio between the Sun and the giant planets would be too small, in which case the potential would no longer be near-Keplerian.
If we were to simulate a more realistic disk mass, the simulations would take proportionally longer and we would need to use fewer particles. Scattering interactions between disk particles and the planets would naturally depopulate the disk, further reducing numerical resolution of the simulation. The instability cannot be captured at low particle numbers \citep{Madigan2018}.  

Our solution \deleted{to this problem }is to model the Sun and the giant planets with a multipole expansion\deleted{. For this, we} keeping only the two largest terms\deleted{ in the series expansion}, the monopole and quadrupole term (the dipole term is zero in the center of mass frame).\deleted{ The monopole term is the standard $\nicefrac{1}{r}$ Keplerian potential with the mass given by the total mass of the system. Because the Sun is much more massive than the sum of the masses of the planets} We ignore the contributions of the planets to the monopole term \added{because this results in a negligible (1 part in thousand) change in the Sun to disk mass ratio}.
In spherical coordinates ($r$, $\theta$, $\phi$), the multipole expansion potential is,
\begin{equation}
    \label{eq:multipole}
\Phi(r, \theta) = - \frac{GM}{r} \left( 1 - \frac{J_2R^2}{r^2} \, P_2\left(\cos{\theta}\right) \right)
\end{equation}
where $J_2$ is a weighting factor for the quadrupole moment, $R$ is the mean radius of the mass distribution, and $P_2\left(\cos{\theta}\right)$ is the $n=2$ Legendre polynomial. 
The first term in the parentheses is the monopole term and the second is the quadrupole term. 
For the giant planets, the orbit-averaged quadrupole moment is given by,
\begin{equation}
    \label{eq:J2iter}
J_2 = \frac{1}{2M_\odot R^2} \sum_{i=1}^4 m_i a_i^2
\end{equation}
where $i$ iterates over the giant planets \citep{Batygin2016}. We further assume that the Sun's inherent $J_2$ moment is negligible compared to the contributions of the giant planets. 
\deleted{Therefore, we emulate Jupiter, Saturn, Uranus, and Neptune (JSUN) as an artificially large $J_2$ moment on the Sun.}

Equation~\ref{eq:multipole} is not a general multipole expansion; we have already implicitly assumed there is no longitudinal ($\phi$) dependence in $\Phi$.
Thus, this expansion assumes the giant planet's orbits have no inclination or eccentricity and their mass is spread out along their orbit. 
\deleted{The $R^2$ term here cancels with the $R^2$ in the potential, thus, the particular value of $R$ has no effect on the value of the quadrupole potential. }
Formally, a multipole expansion only converges to the actual potential for $r > d$ where $d$ is the size of the system. 
Thus, the \replaced{rigorous}{mathematically correct} method would be to set $R=a_N$, the semi-major axis of Neptune, and remove any particles that went inside $R$. 

Due to the artificially strong self-stirring in our low-$N$, large-mass disks, most bodies in our simulations violate this requirement during integration, and, if we removed them, we would end up with a depopulated disk that is numerically unable to undergo the instability.
Therefore, we ignore this convergence requirement.
\deleted{Such self-stirring isn't expected to happen in the outer solar system, and we are purely interested in how an external source of orbital precession affects the instability and not interested in other dynamical effects of the giant planets on the instability \citep[e.g. scattering, see][]{Fleisig2020}.

Thus, the analysis here is not meant to perfectly replicate the solar system. Instead, it is designed to specifically study the affect of an external source of precession on the inclination instability. 
These results can then be applied to the solar system to determine if the giant planets induce enough differential apsidal {precession} in the orbits of the bodies beyond Neptune \replaced{such that the inclination instability could not occur there}{to suppress the inclination instability}.}

We do not use the actual $J_2$ value of the giant planets because of the unrealistic disk mass and $N$ used in our simulations.
For a given set of simulations, we fix the number of particles, $N$ and the mass of the disk, $\Md$, and vary the $J_2$ value until we find the instability is suppressed.\deleted{ It requires numerous simulations to find this $J_2$ value.}
To extrapolate our results to the solar system, we determine how \replaced{they}{these $J_2$ values} scale with $N$, $\Md$, and orbital configuration of the disk.  

\subsection{Initial Orbit configurations}

In this paper we simulate two distinct systems, a compact configuration and a scattered disk configuration. 
The compact configuration is a thin, mono-energetic disk of orbits (nearly identical semi-major axes), and the scattered disk configuration models a population of bodies with equal perihelion and an order-of-magnitude range in semi-major axis. 

In the compact configuration, the disk of orbits is initialized to have a semi-major axis $a$ distribution drawn uniformly in \replaced{$[0.9,1.1)$ ~sim.\ units}{$[0.9,1.1]$}, eccentricity $e=0.7$, and inclination $i=10^{-4}\,{\rm rad}$. The disk is initially axisymmetric ($\omega$ and $\Omega$ and mean anomaly, $\mathcal{M}$, drawn from a uniform distribution in \replaced{$[0,2\pi)$}{$[0,2\pi]$}). The total mass of the disk is $10^{-3}\,M$ and the number of disk particles, $N = 400$.
\added{This configuration is ideal for physical analysis.}

\deleted{In our simulations, we exploit the scale-free nature of the Newtonian $N$-body problem. 
We use $a = 1$ as our unit simulation radius, converting to $a\sim100$~AU or $a\sim250~AU$ when necessary by scaling the Keplerian orbital period.}

In the scattered disk configuration, orbits are initialized with an order-of-magnitude range in semi-major axes and identical perihelion distances.
Specifically, we draw the orbit's semi-major axis from an $a^{-1}$ distribution in the range \replaced{$[1,10)$ sim.\ units}{$[1,10]$}, define eccentricity $e$ from the relation $e = 1 - \nicefrac{p}{a}$ for \replaced{some}{a} chosen perihelion $p$, and draw inclination $i$ from a Rayleigh distribution with \replaced{$\mu_i = 5^\circ$}{a mean inclination of $5^\circ$}.\footnote{The instability can occur in scattered disk simulations with initial inclinations drawn from Rayleigh distributions with means up to $\sim15^\circ$, but it's hard to measure the instability growth rate in these systems because the instability is linear for a short time.} The disk is initially axisymmetric with $\mathcal{M}$ drawn uniformly in the range \replaced{$[0,2\pi)$}{$[0,2\pi]$}.

We look at two different scattered disk configurations: `sd100' and `sd250'. 
The `sd100' configuration represents a scattered disk with inner-most semi-major axis of 100 AU and a perihelion of 30 AU, while the `sd250' configuration represents a scattered disk with the same perihelion but an inner-most semi-major axis of 250 AU.

\deleted{The only initial difference in the two simulations are the perihelions, which are $p=0.3$ and $p=0.12$ in simulation units respectively.}
We run these two simulations to explore the effect of distributing the peak of the mass density of the scattered disk in a different location. Apsidal precession due to the $J_2$ moment is a steep function of semi-major axis, $a^{-7/2}$; perhaps the gravitational torques between orbits in a scattered disk with peak mass density at larger radius can better resist the differential precession from the giant planets?

\added{The Newtonian $N$-body problem is scale-free. Simulation times are presented in units of the secular timescale, 
\begin{align}
	t_{\rm sec} \sim \frac{1} {2\pi} \frac{M}{\Md} P,
	\label{eq:sec-time}
\end{align}
where $P$ is the orbital period at $a=1$. 
In this paper, $M_d = 10^{-3}\,M$ such that $t_{\rm sec} \approx 160\,P$.
$a=1$ may be scaled to, for example, $a=100$ AU with the conversion $P = 1000$ yr. 
The $J_2$ potential is not scale-free however. In simulations with added $J_2$ we appropriately scale the semi-major axes, $a_i$, in equation~\ref{eq:J2iter}, and present our results in solar system units (distances in AU etc.).}

\section{J2 and the Inclination Instability}
\label{sec:suppress}

\begin{figure}[!tb]
    \centering
    \includegraphics[width=\columnwidth]{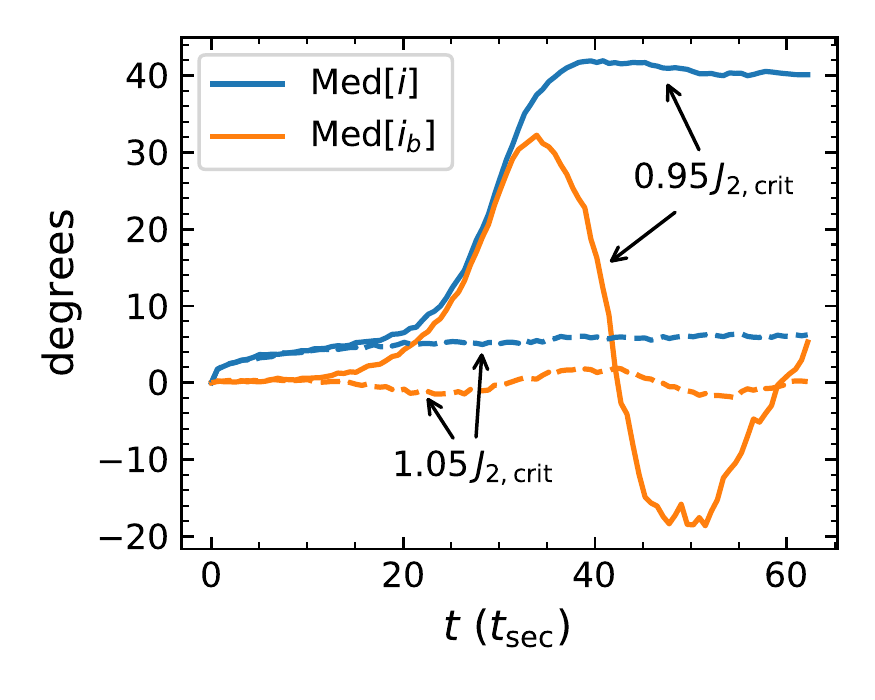}
    \caption{Median inclination (blue) and $i_b$ (orange) of disk orbits for two different simulations, one with $J_2$ less than $\Jc$ (solid) and one with $J_2$ greater than $\Jc$ (dashed), vs. time measured in secular times ($t_{\rm sec} \approx 160\, P$). The simulation with added $J_2$ less than $\Jc$ is susceptible to the inclination instability while the other is not. The two simulations have very similar $J_2$ values, only different by about 20\%, showing the abruptness of the transition from unstable to stable.}
    \label{fig:on-off}
\end{figure}

The inclination instability timescale, $\te$, scales \added{linearly} with the secular time. 
\replaced{In addition to this, $\te$}{It also} depends \added{non-trivially} on $N$ and orbital configuration.
We use the orbital angle coordinates defined in \citet{Madigan2016} to describe the instability and quantify its timescale. The angles represent rotations of the orbit about its semi-major (${\hat a}$) axis, semi-minor (${\hat b} \equiv \hat{j} \times \hat{a}$) axis and angular momentum vector ($\hat{j}$), respectively,
\begin{subequations}
\begin{align}
  \ia &= \arctan\left[\unitvectorslope{b}\right], \\
  \ib &= \arctan\left[-\unitvectorslope{a}\right], \\
  \ie &= \arctan\left[\hat{a}_{\text{y}},
    \hat{a}_{\text{x}}\right].
\end{align}
\label{eq:iaibie}
\end{subequations}
The subscripts $x$, $y$, and $z$ denote an inertial Cartesian reference frame with unit vectors, $\hat{x}$, $\hat{y}$, and $\hat{z}$. 
These angular coordinates are useful for understanding the effect of torques on orbits.

The inclination instability is characterized by exponential growth in median $\ia$ and $\ib$ with opposite sign (i.e.\ if $\ia$ increases to positive values, $\ib$ increases to negative values).
A constant ratio $\ib/\ia$ implies a constant angle of perihelion, as for small inclinations $\omega(\ia,\ib) \sim \arctan{ |\ib/\ia|}$ ($+\pi$ if \ia $< 0$). 
We use the exponential growth of median $\ib$ as a diagnostic for the instability and define the inverse of its growth rate, $\gamma$, as $\te$. 
As orbits incline, their eccentricities decrease. This means that the magnitude of the angular momentum vectors of all orbits increase (semi-major axes remain constant apart from scatterings due to two-body relaxation). This may seem counter-intuitive at first, but the {\it vector} sum of all the angular momenta is conserved.  

\subsection{Compact Configuration}

\begin{figure}[!tb]
    \centering
    \includegraphics[width=\columnwidth]{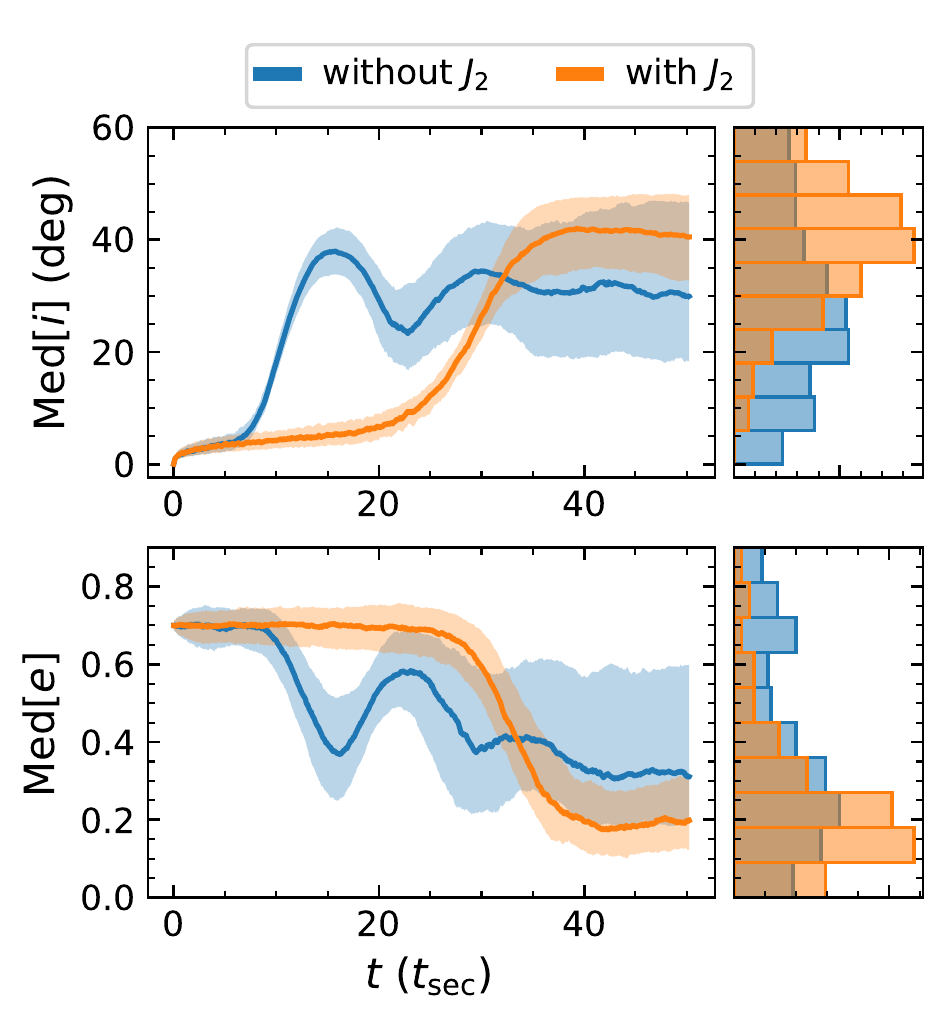}
    \caption{Median inclination (deg) and eccentricity for two simulations, one with added $J_2$ and one without $J_2$. The ``with $J_2$'' simulation has an added $J_2$ in the transition region ($J_2 = 0.9\, \Jc$). The left column shows the time evolution of the median orbital elements with their upper and lower quartiles. The right column is a histogram showing the distribution of the particle's orbital elements in the simulation at the end of the plotted time evolution ($\sim50\,t_{\rm sec}$). The ``with $J_2$'' simulation has a smaller growth rate than the ``without $J_2$'' simulation, but it reaches a higher median inclination and lower median eccentricity post-instability. }
    \label{fig:pi-orb-elem}
\end{figure}

A sufficiently large J2 value suppresses the inclination instability. We call the threshold value above which the disk does not undergo the instability $\Jc$. 
\deleted{In section~\ref{sec:scaling} we discuss the numerical value of $\Jc$ and its physical meaning.}
For $J_2 < \Jc$, we find two different regimes:
\begin{enumerate}
    \item The `instability-dominated region' defined by $J_2 \leq 0.1\,\Jc$. Here the system is unaffected by the additional $J_2$.
    \item The ``transition region'' defined by $(0.1 \-- 1) \Jc$. Here the dynamics of the instability are altered by the presence of the $J_2$, but the instability still occurs.
\end{enumerate}

In Figure~\ref{fig:on-off}, we plot the median inclination $i$ and $\ib$ of a disk of particles in two simulations, one with $J_2 = 0.9~\Jc$ and another with $J_2 = 1.1~\Jc$. 
This figure shows that the inclination instability is suppressed for $J_2 > \Jc$ and that the transition around $\Jc$ is rapid, with the inclination behavior of the disk changing dramatically for only slight changes ($\sim20\%$) in the value of added $J_2$.

In Figure~\ref{fig:pi-orb-elem}, we show that the average post-instability orbital elements of the disk are different in the transition region ($J_2 = 0.9 \, \Jc$). 
The orbits attain higher (lower) post-instability inclinations (eccentricities) on average than systems with no/low $J_2$, despite the fact that the instability growth rate is reduced by the added $J_2$. The right columns show histograms of the orbital elements at \replaced{$8000$ orbital periods}{$50$ secular times}. The histograms are limited in range for clarity; two out of eight hundred bodies have reached \added{polar orientations of} $i \gtrsim 90^\circ$. 

\begin{figure}[!tb]
    \centering
    \includegraphics[width=\columnwidth]{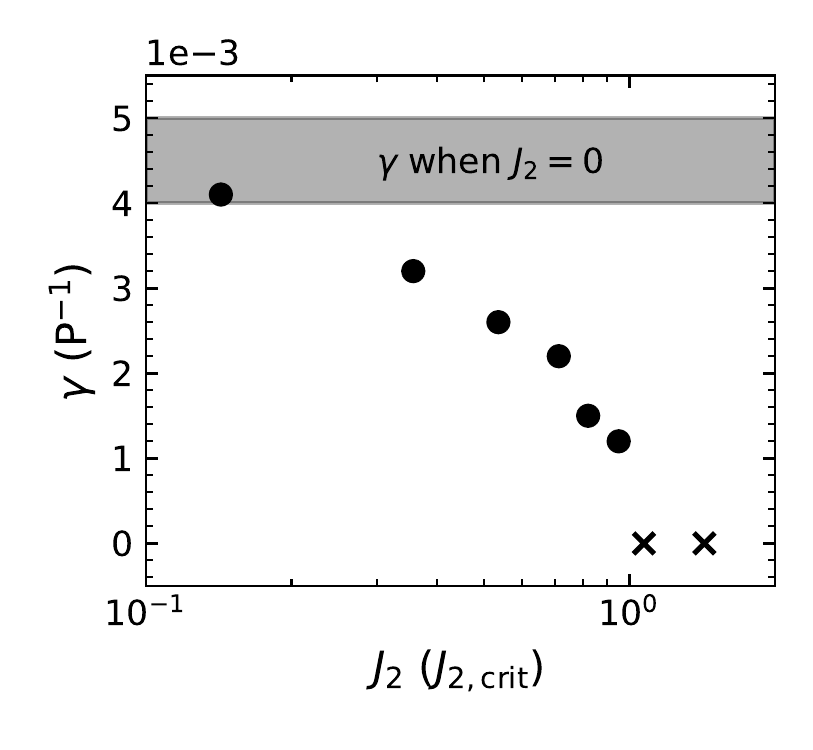}
    \caption{Growth rate of the inclination instability ($\gamma$) as a function of added $J_2$ moment. For $J_2 \lesssim 0.1 \Jc$ the growth rate of the instability is the same as if there were no added $J_2$. In the region between 0.1 to 1.0 $\Jc$ the growth rate steadily drops until the instability disappears for $J_2 > \Jc$. Above this, the disk is stable and the growth rate becomes imaginary as signified by the change in marker.}
    \label{fig:gr-change}
\end{figure}

In Figure~\ref{fig:gr-change}, we show that the growth rate of the instability decreases across the transition region. 
At $\sim 0.1 \, \Jc$, the growth rate of the instability is identical to the instability with no $J_2$ moment, and at $\Jc$, the instability has a growth rate of zero.
Above $\Jc$, we find that the median $i_b$ of the disk oscillates rather than grows exponentially; the growth rate is imaginary. 

\begin{figure}[!tb]
    \centering
    \includegraphics{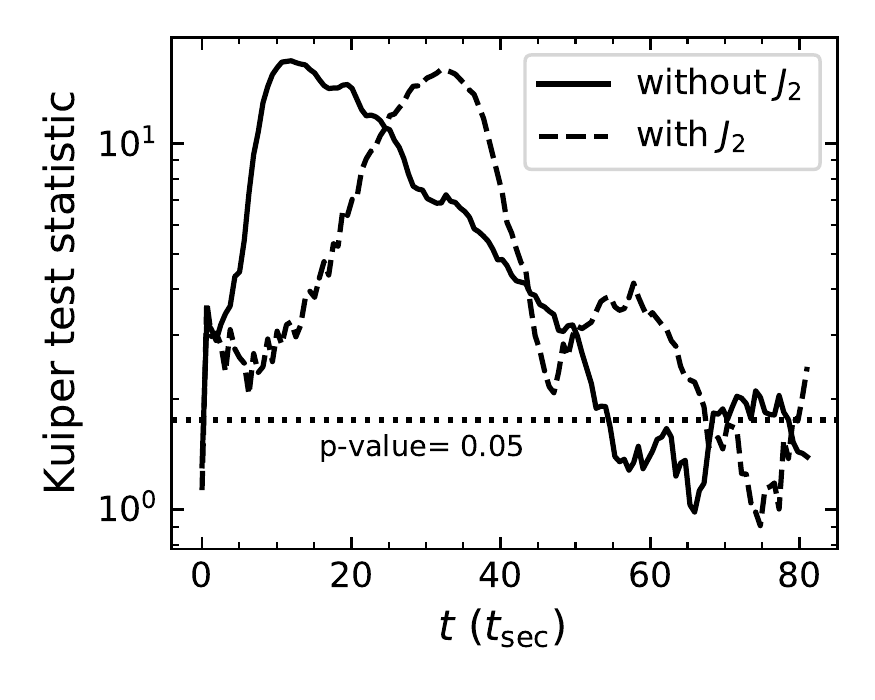}
    \caption{Kuiper's test statistic comparing the $\omega$ distribution of $N$-body simulations to a uniform distribution as a function of time. Two simulations are shown, one without $J_2$ and another with added $J_2$ in the transition region ($J_2 = 0.8\Jc$). \replaced{Horizontal lines mark}{A horizontal line marks} the test statistic \replaced{values for critical $p$-values}{value for a $p$-value of 0.05}. The test statistic reflects the dynamical behavior of $\omega$ over the course of the simulation with the obvious peak corresponding to the peak clustering during the instability.}
    \label{fig:w-cluster}
\end{figure}

\begin{figure*}
    \centering
    \includegraphics[]{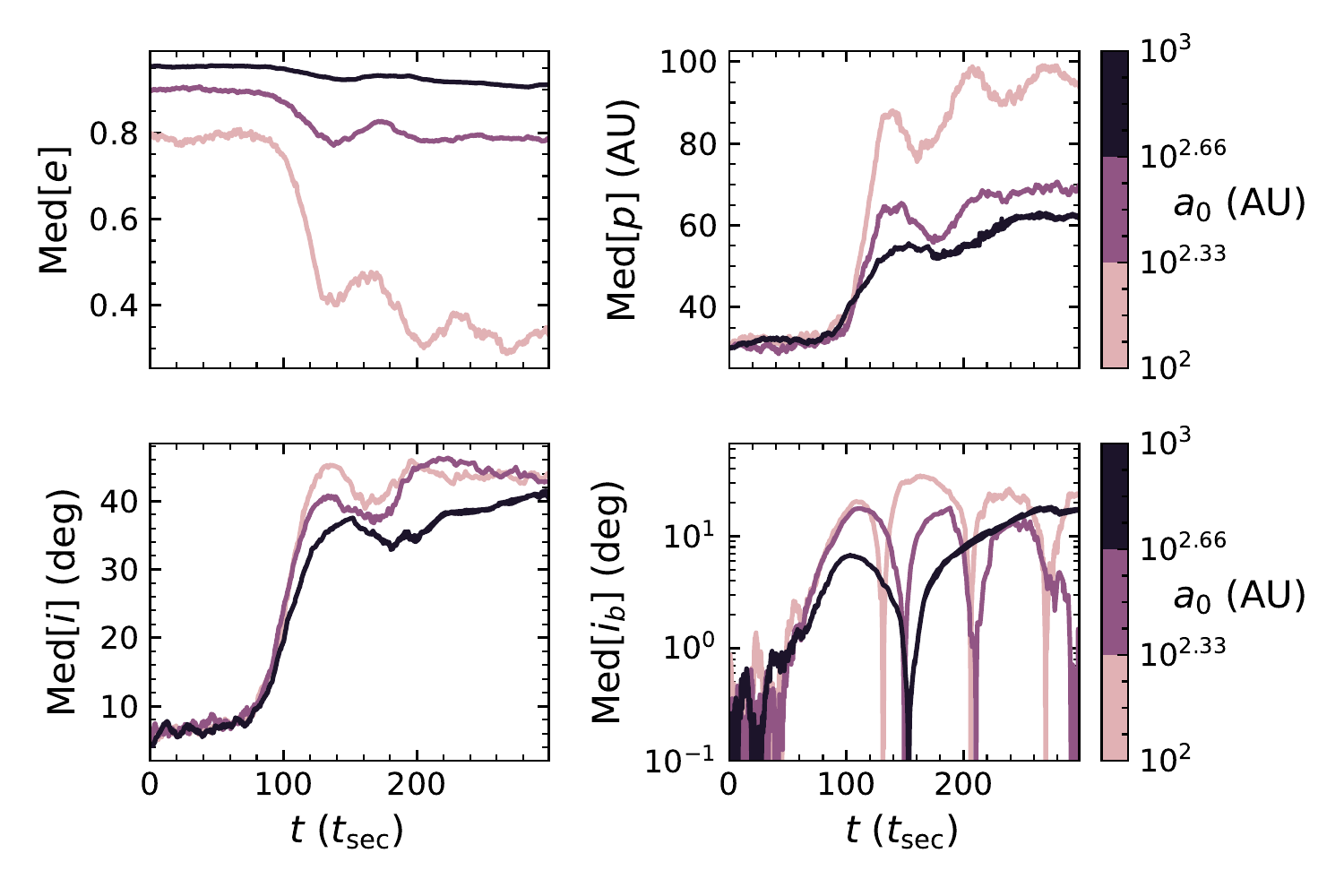}
    \caption{Median eccentricity (top left), perihelion distance (top right), inclination (bottom left) and $\ib$ (bottom right) as a function of time in units of secular times for a scattered disk (`sd100') simulation with no $J_2$ moment. Orbits have been binned by their initial semi-major axis. The lower the initial eccentricity/semi-major axis, larger the change in eccentricity/perihelion during the instability. All bodies attain a similar post-instability inclination, with the larger semi-major axis bodies attaining slightly lower final inclinations. The larger semi-major axis orbits have slightly lower instability growth rates (slope of Med[$i_b$] during the instability) than smaller semi-major axis orbits, but the instability begins at the same time at all radii.}
    \label{fig:insta-sd}
\end{figure*}

In Figure~\ref{fig:w-cluster}, we show the effects of $J_2$ on the clustering of argument of perihelion, $\omega$, \added{in the compact configuration }using the Kuiper test, a variation of the Kolmogorov-Smirnov test that is applicable to circular quantities \citep{Kuiper1960}.
We use the test to compare the simulation $\omega$ distribution to a uniform distribution for two different simulations, one with $J_2$ in the transition region and one with no added $J_2$.
We show the test statistic value for a $p$-value of 0.05.
Larger test statistic values correspond a greater likelihood that the simulation $\omega$ distribution is not uniform.
In both simulations, the test statistic is initially consistent with a uniform distribution.
Within a single orbit, the system develops a bi-modal distribution in $\omega$ with peaks at $0^\circ$ and $180^\circ$ due to small oscillations in $i_a$.
Later, the test statistic increases to a large peak as the instability clusters the orbit's $\omega$. 
Post-instability, the $\omega$-clustering is not maintained, and differential precession washes out the clustering.

Surprisingly, the duration of $\omega$-clustering isn't significantly changed in the transition region. 
The $J_2$ potential term causes prograde ($\dot{\omega} > 0$) precession, and the {\it post-instability} disk potential causes retrograde precession.
One might expect that the two competing sources of precession would reduce the overall precession rate, and increase the duration of $\omega$-clustering.
However, this is not what we see.
When $J_2$ is added to the system, the growth rate slows, and the rise time to peak clustering increases. The mean $\omega$ precession rate decreases, but the {\it differential} precession rate increases.
Thus, the $\omega$-clustering is washed out faster.
Overall, the duration of clustering is relatively unchanged \added{in the compact configuration}.

In summary, we find that the addition of a $J_2$ term to the Keplerian potential suppresses the inclination instability above a critical value, $\Jc$. As this critical value is approached from below, the post-instability orbital elements and growth rate of the instability are changed in a transition region, 0.1\,$\Jc$ to $\Jc$. Finally, we find that the duration of $\omega$-clustering is unchanged in this transition region.

\subsection{Scattered Disk Configuration}
\label{sec:scat-disk}

In previous publications, we focused on the compact configuration for ease of analysis.
However, we have explored the inclination instability in a range of different orbital initial conditions. 
Our findings can succinctly be summarized: compact systems with mean eccentricity $\gtrsim 0.5$ and/or mean inclinations $\lesssim 20^\circ$ are unstable, systems with an order of magnitude spread in semi-major axis {\it with constant eccentricity} are either stable or have very small growth rates, and systems with an order of magnitude spread in $a$, but $\nicefrac{de}{da} > 0$, i.e. the scattered disk, are unstable.

In Figure~\ref{fig:insta-sd}, we show the inclination instability in a system with scattered disk initial conditions (`sd100') and no $J_2$ moment. 
The disk undergoes the instability simultaneously at all radii, though the orbits at larger semi-major axis have a slightly smaller growth rate.
In general, the lower the initial eccentricity (and semi-major axis) of the orbit, the larger the change in eccentricity during the instability and the larger the final perihelion distance.
The final median inclination is similar for all semi-major axis bins ($i \approx 40^\circ$).

\begin{figure*} 
    \centering
    \includegraphics[]{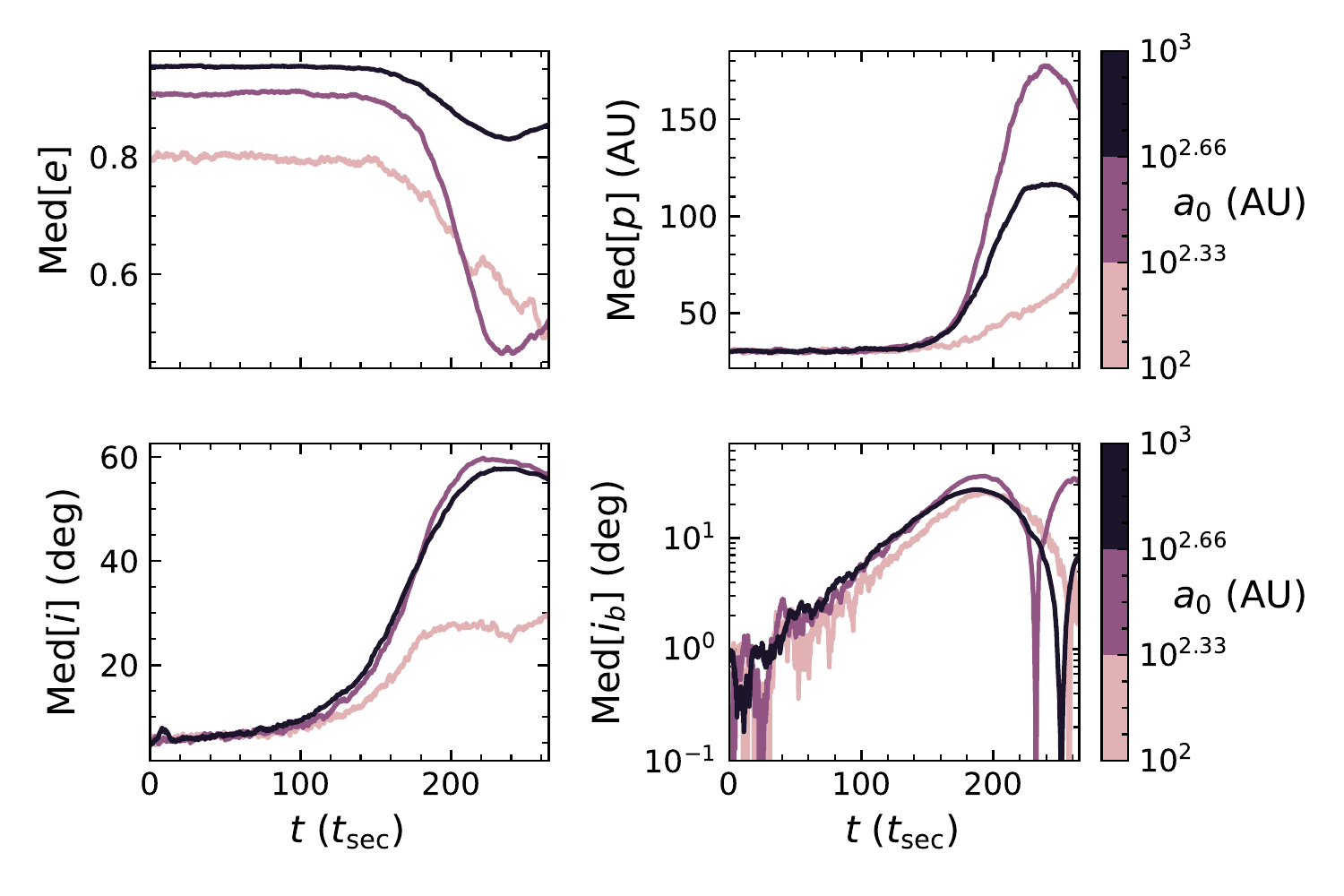}
    \caption{Median eccentricity (top left), perihelion distance (top right), inclination (bottom left) and $\ib$ (bottom right) as a function of time for a scattered disk (`sd100') orbital configuration with $J_2$ in the transition region. The orbits have been binned by their initial semi-major axis. Compared to Figure~\ref{fig:insta-sd}, the growth rate of the instability is smaller and all bins have the same growth rate. The inner-most semi-major axis bin is barely unstable (with a reduced post-instability inclination), and the outer bins have larger post-instability perihelia and larger post-instability inclinations.}
    \label{fig:insta-sd-J2}
\end{figure*}

Figure~\ref{fig:insta-sd-J2} shows the same information as Figure~\ref{fig:insta-sd}, but for a simulation with added $J_2$ in the transition region (in this case $J_2 \sim 0.9\,\Jc$). 
Again, the instability occurs simultaneously throughout the disk. 
The smallest semi-major axis bin is barely unstable, however, and has a lower post-instability inclination and a higher eccentricity. 
This is due to the significant differential apsis precession caused by the added $J_2$. 
The larger semi-major axis bins have larger post-instability inclinations ($i \approx 60^\circ$), lower eccentricities ($0.5 \lesssim e \lesssim 0.85$), and larger perihelia ($100\,{\rm AU} \lesssim p \lesssim 150\,{\rm AU}$) than they do in simulations without $J_2$. 

Overall, the addition of the $J_2$ moment to simulations has a similar effect on the scattered disk orbital configuration as it has on the compact configuration, i.e., increased (decreased) post-instability inclination (eccentricity) and reduced instability growth rate. 
One significant difference is the inner-most part of the disk barely undergoes the instability. 
Indeed, if we simulate the inner portion of the disk ($a \in [100,200]\,{\rm AU}$) without the outer portion it does not undergo the instability at all.
The inner portion of the disk is being pulled along by the outer portion as the outer portion undergoes the instability. 
We can think of this as the disk having two components, a stable component and an unstable component.
The inner-most part is stabilized by differential precession from the $J_2$ moment while the outer portion is still unstable ($J_2$ precession has a steep $a^{-7/2}$ dependence). 
Below $\Jc$, the inner-most component is small enough that it can be coerced into instability by the outer-most portion.
At the critical $J_2$, the \added{stable, }inner-most component of the disk is massive enough that the outer portion of the disk is held back from lifting out of the plane.
The inclination instability is a \replaced{{\it global}}{global} phenomenon, and we find that the disk as a whole is stabilized if $\sim30\%$ of the mass is in the stable component.

\begin{figure*}[!htb]
    \centering
    \includegraphics[]{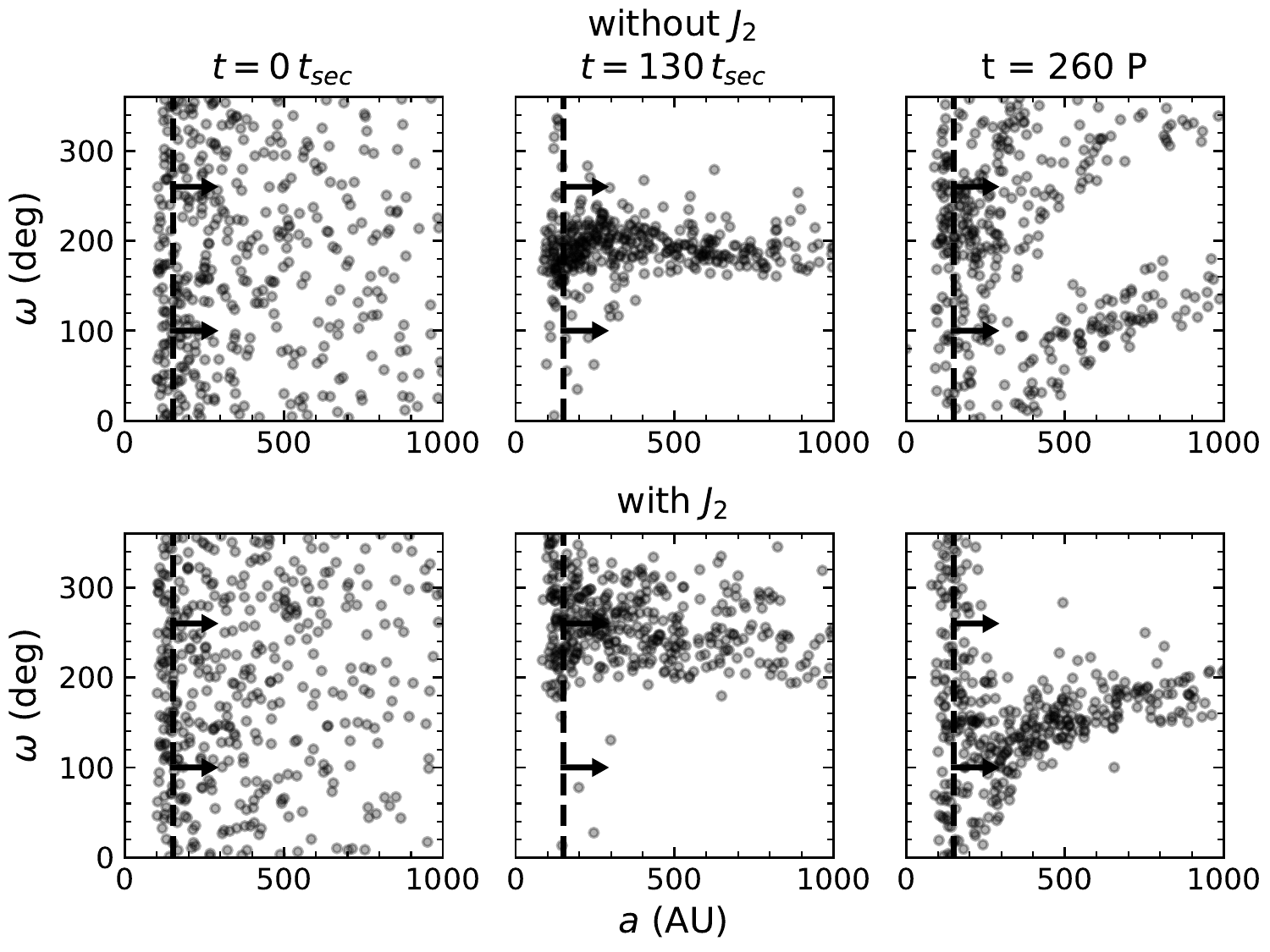}
    \caption{Argument of perihelion, $\omega$, as a function of semi-major axis, $a$, for a simulation without $J_2$ (top) and one  with added $J_2$ in the transition region (bottom), corresponding to Figures \ref{fig:insta-sd} and \ref{fig:insta-sd-J2}. The vertical dashed line marks $a = 150\,{\rm AU}$ \deleted{, which using the conversion to solar system units $a = 1 = 100\,{\rm AU}$ corresponds to $150\,{\rm AU}$}. 
    (Left column) orbits initially have a uniform random $\omega$ distribution. 
    (Middle column) $\omega$ is clustered while orbits undergo the inclination instability. 
    (Right column) $\omega$-clustering is lost in the simulation without $J_2$ due to differential precession, but it is maintained for \replaced{$a \gtrsim 2$}{$a \gtrsim 200\,{\rm AU}$} in the simulation with $J_2$ due to the reduced differential precession and instability growth rate.}
    \label{fig:omega-cluster-sd}
\end{figure*}

\added{
We find that the duration of $\omega$-clustering in the scattered disk simulations is enhanced by the addition of $J_2$ in the transition region, in contrast to our findings for the compact configuration.
In Figure~\ref{fig:omega-cluster-sd}, we plot the argument of perihelion, $\omega$, as a function of semi-major axis, $a$, for each disk orbit at three different times corresponding to the simulations shown in Figures \ref{fig:insta-sd} and \ref{fig:insta-sd-J2}. 
At the beginning of each simulation, the distribution of $\omega$ values is uniform. 
Later, the instability causes $\omega$ values to cluster. 
After $\sim260\, t_{\rm sec}$, the $\omega$ values for $a \gtrsim 200\,{\rm AU}$ are significantly less clustered in the simulation without $J_2$ while the simulation with $J_2$ still retains significant $\omega$-clustering.
This difference is due to both the reduced growth rate (and delayed instability saturation time) of the instability due to the added $J_2$ potential and the reduced differential precession rate in the $a \gtrsim 200\,{\rm AU}$ portion of the disk due to competition between the disk and $J_2$ precession.
}

The global nature of the instability has an interesting consequence on the perihelion distribution of the post-instability orbits.  
As we see in Figure~\ref{fig:insta-sd} in which all the orbits are unstable, orbits of different semi-major axes end up with similar mean inclinations. Specific orbital angular momentum increases with semi-major axis across the scattered disk ($\sim10\%$ change from 100 to 1000 AU). This means that as orbits incline, those at lower semi-major axis will gain a larger fractional increase in orbital angular momentum than those at higher semi-major axis. This results in orbits at lower semi-major axis decreasing their eccentricities and increasing their perihelia more so than those at higher semi-major axis. This naturally generates a perihelion gap at the inner edge of the disk that has undergone the instability. 

\begin{figure*}[!htb]
    \centering
    \includegraphics[]{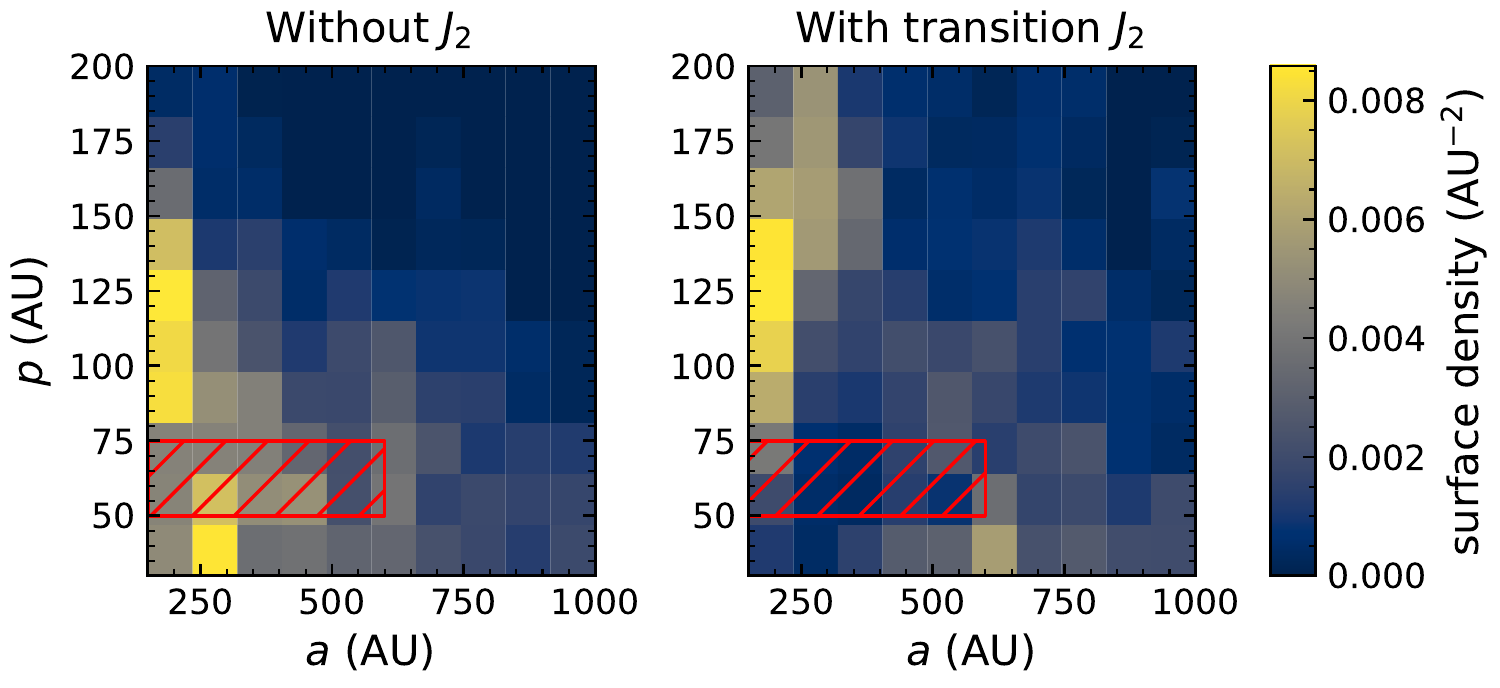}
    \caption{Time-averaged surface density in the semi-major axis ($a$) - perihelion ($p$) plane for two scattered disk (`sd100') simulations, one without $J_2$ and one with $J_2$ in the transition region ($J_2 \sim 0.9\,\Jc$). For this plot, we histogram the $a$ and $p$ of each particle in the simulation at all times post-instability. Areas of high density will have more observable bodies than areas of low density. Also drawn is a red box covering the observed ``perihelion gap'' in the solar system \deleted{, using the conversion $a = 1 = 100$ AU}. When $J_2$ is added to the simulations in the transition region, the region of $a$-$p$ space corresponding to the observed perihelion gap in the solar system is vacated by the inclination instability. The size of the gap changes with the added $J_2$; larger $J_2$ values produce a larger gap.}
    \label{fig:pericenter-gap}
\end{figure*}

In Figure~\ref{fig:pericenter-gap}, we show a 2D histogram of time-averaged post-instability values of semi-major axis vs perihelion for the two simulations shown in Figures~\ref{fig:insta-sd} and~\ref{fig:insta-sd-J2} (scattered disk configurations without $J_2$ and with $J_2$). We use a time-average to get sufficient numerical resolution to make this plot.
\deleted{Using the conversion $a = 1 = 100$ AU,} We've added a red box to the figures to show the observed perihelion gap between VP$_{113}$ and Sedna and the rest of the minor bodies (see Figures 1 and 2 in \citet{Kavelaars2020}).
In the simulation with $J_2$, the inclination instability empties the region corresponding the observed perihelion gap.
The size of the region vacated by the inclination instability is related to the magnitude of $J_2$, a larger $J_2$ vacates a larger region of $a$-$p$ space.

\section{Scaling to the Solar System}
\label{sec:scaling}

Our goal in this section is to explain how $\Jc$ depends on system parameters such as number of particles, $N$, mass of disk, $\Md$, and initial orbital configuration, which allows us to then extrapolate our simulation results to the solar system.

The instability mechanism relies on a secular average where the individual bodies can be approximated as rings in the shape of the body's osculating Keplerian orbit with a linear mass density inversely proportional to \replaced{the velocity of the body at that point on its orbit}{its instantaneous velocity}. 
The validity of this average depends on how quickly the body's osculating Keplerian orbital elements change.
\deleted{The slower the osculating orbit changes, the better the approximation works, and the stronger the instability is.}
If the osculating orbit changes rapidly the approximation fails and the mutual secular torques responsible for the instability weaken to the point that the instability can no longer occur.
\deleted{In particular, rapid differential apsidal precession can cause this approximation to fail.}
As the instability relies on \replaced{{\it mutual}}{inter-orbit} secular torques, it is \replaced{{\it mutual}}{inter-orbit} or differential apsidal precession that is responsible for the weakening of the secular torques.
\deleted{Thus, the magnitude of the {\it differential} apsidal precession within the disk determines the suppression of the instability not the absolute magnitude of the apsidal precession.}
\deleted{
As we will show, } The addition of the quadrupole term to the potential increases differential apsidal precession within the disk.
However, the disk itself also causes apsidal precession in its constituent orbits, and this source of apsidal precession must be considered in combination with that from the $J_2$. 

\begin{figure}
    \centering
    \includegraphics[width=\columnwidth]{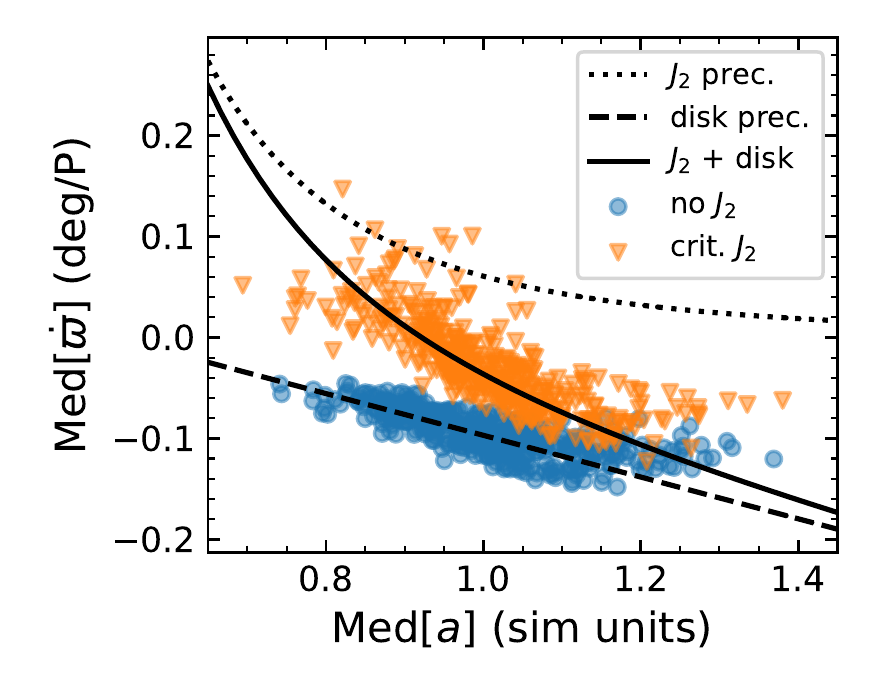}
    \caption{
    Median apsidal precession rate as a function of median semi-major axis (calculated from the first \replaced{1000 orbits}{$\sim6$ secular times}) of a simulation with a compact configuration. Orange triangles show apsidal precession rates from a simulation with $J_2 \approx \Jc$, and blue dots show one without $J_2$. 
    The lines on the plot show models for the different precession sources. The dotted line shows the $J_2$ contribution to $\dot{\varpi}$, the dashed shows the disk contribution, and the solid line shows the sum of the dashed and dotted lines. The differential precession rate (slope of solid line) is enhanced by the presence of the added $J_2$ although the average precession rate is reduced.}
    \label{fig:prec-model}
\end{figure}

The addition of the quadrupole term causes secular changes in the $\omega$ and $\Omega$ of the orbits in the disk. 
Assuming the added quadrupole term is a small perturbation on the \deleted{$\nicefrac{1}{r}$} potential of the central body, the evolution of the osculating Keplerian elements of orbits in the potential can be determined with the disturbing function formalism and the Lagrange planetary equations,
\begin{subequations}
    \begin{align}
        \dot{\omega} &= \frac{3J_2}{4} \, n \, \frac{R^2}{a^2} \, \frac{5\cos^2{i} -1}{\left(1 - e^2\right)^2}, \\
        \dot{\Omega} &= -\frac{3J_2}{2} \, n \, \frac{R^2}{a^2} \, \frac{\cos{i}}{\left(1 - e^2\right)^2},
    \end{align}%
    \label{eq:omegas}%
\end{subequations}
where $n$ is the mean motion of the body, $n^2 a^3 = \mu = GM$.\deleted{(for completeness we include the derivation in appendix~\ref{app:J2}). The disk orbits are initialized at low inclination.}
$\varpi = \omega + \Omega$ gives the apsidal angle of the orbit and $\dot{\varpi}$ the apsidal precession rate. 
Therefore, the apsidal precession rate from the added $J_2$ is,
\begin{equation}
\label{eq:J2prec}
\dot{\varpi}_{J2} = \frac{3J_2}{4}\,n\,\frac{R^2}{a^2} \frac{5\cos^2{i} - 2\cos{i} - 1}{\left(1-e^2 \right)^2}. 
\end{equation}
For $i \lesssim 46^\circ$, which holds for our initial disk configurations, apsis precession due to $J_2$ is prograde (with respect to orbital motion).
\replaced{Our simulation units give $G= M = R = 1$. Thus, the apsidal precession becomes, 
\begin{equation}
    \dot{\varpi}_{J2} = 270\,J_2 \, a^{-\nicefrac{7}{2}} \, \frac{5\cos^2{i} - 2\cos{i} - 1}{(1-e^2)^2}
\end{equation}
with $\dot{\varpi}_{J2}$ given in degrees per orbit at $a=1$. 
For the compact configuration, $i \approx 0^\circ$ and $e\approx0.7$ yielding,
\begin{equation}
    \label{eq:J2prec-compact}
    \dot{\varpi}_{J2} = 2076\,J_2 \, a^{-\nicefrac{7}{2}},
\end{equation}
where $a$ and $J_2$ are given in simulation units and $\dot{\varpi}_{J2}$ in deg per orbit. 
This $J_2$ contribution to apsidal precession is shown as a dotted line in Figure~\ref{fig:prec-model}. 
For reference, $\Jc = 2.6 \times 10^{-5}$ for this simulation.}{In the compact configuration, $e\approx0.7$ and $i\approx0$. In Figure~\ref{fig:prec-model}, we show Equation~\ref{eq:J2prec} with these approximations as a dotted line. For reference, \replaced{$\Jc = 2.6 \times 10^{-5}$}{$\Jc = 0.26$} for this simulation (length scaled to 100 AU).}

The disk potential induces retrograde precession (see Appendix~\ref{app:disk} for differences between orbital configurations) whereas added $J_2$ potential induces prograde precession. \deleted{One might therefore assume that, in combination, the differential precession rate will be reduced, and the time scale over which secular torques act lengthened by the addition of the quadrupole potential. 
This is partly correct.} The two precession sources \deleted{do} compete and the \replaced{{\it absolute}}{mean} precession rate is reduced. However, secular torques are not always strengthened by the added $J_2$ potential. 
In the compact configuration, $|\dot{\varpi}_{J2}|$ decreases with semi-major axis and the precession rate due to the disk, $|\dot{\varpi}_d|$, increases with semi-major axis.  
The result, as shown in Figure~\ref{fig:prec-model} with a solid line, is an amplified differential precession rate (slope).
Thus, in this orbital configuration, the added $J_2$ weakens the gravitational torques between orbits which hinders the growth of the instability.
On the other hand, in the scattered disk configurations $|\dot{\varpi}_d|$ decreases with semi-major axis.
Therefore, the scattered disk configuration does a better job of resisting the added $J_2$.
For example, with $N=400$ and $\Md = 10^{-3}\,M$, the scattered disk configuration has \replaced{$\Jc = (3.25\pm 0.25)\times10^{-5}$}{$\Jc = 0.325\pm 0.025$} while the compact configuration has \replaced{$\Jc = (2.58 \pm 0.08)\times10^{-5}$}{$\Jc = 0.258 \pm 0.008$} \added{(length scaled to 100 AU)}.
Despite having a much lower mass density (and growth rate), the scattered disk configuration handles the added $J_2$ better than the compact configuration.
\begin{figure}
    \centering
    \includegraphics[width=\columnwidth]{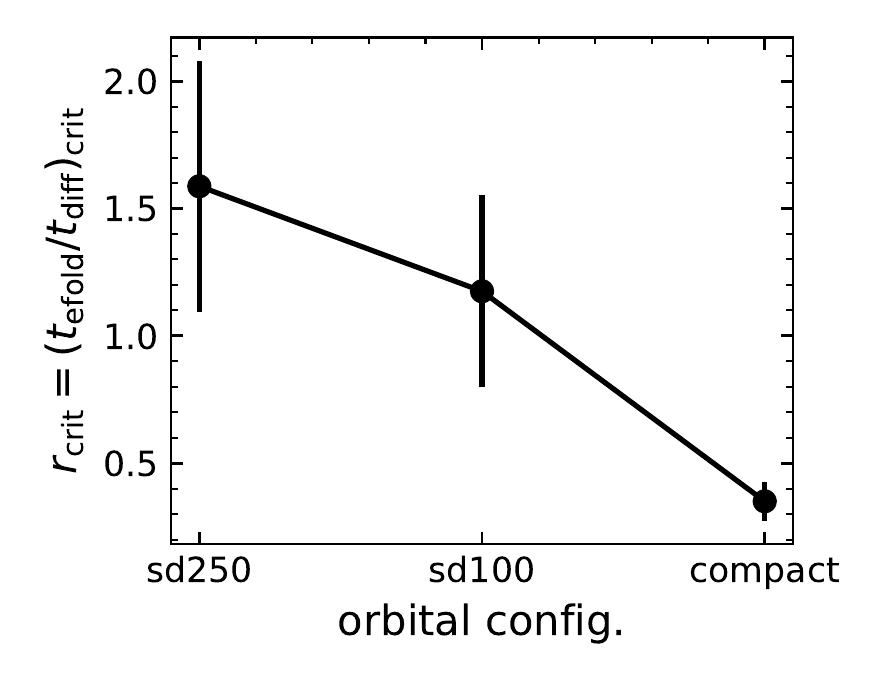}
    \caption{
    Ratio of e-folding to differential apsidal precession timescale at the critical $J_2$ where the instability is suppressed for different simulation initial conditions, $\rc$. 
    $t_{\rm e-fold}$ is measured when $J_2=0$ and $\td$ is measured when $J_2\approx\Jc$.
    The timescales are roughly equal ($\rc\sim1$) at the transition from instability to stability.
    The scattered disk simulations have a larger $\rc$ indicating that these orbital configurations can resist more differential precession relative to their instability growth rate than the compact configuration.
    }
    \label{fig:r-from-sims}
\end{figure}

\deleted{The point where the instability is suppressed depends on the relative strength of the differential apsidal precession and secular torques between orbits. 
We expect the transition from instability to stability in the disk to occur when the strengths of these two dynamical processes are comparable.}
\added{We find that $\Jc$ corresponds to the point where the differential apsidal precession and the inclination instability timescales are comparable.}
The timescale for the instability is defined as the inverse of the exponential growth rate, $t_{\rm e-fold} = \gamma^{-1}$, which we obtain from simulations by fitting the exponential growth of the median $i_b$ of the disk orbits \citep{Madigan2016,Madigan2018}. 
\deleted{Here} We define the differential precession timescale as,
\begin{equation}
    t_{\rm diff} = \frac{1~{\rm rad}}{\dot{\varpi}_{\rm diff}}
\end{equation}
where $\dot{\varpi}_{\rm diff}$ is the total apsidal differential precession rate of the disk. 
We calculate $\dot{\varpi}_{\rm diff}$ from our simulations by calculating the difference in precession rate between the fastest precessing quartile of the disk and the slowest precessing quartile, $\dot{\varpi}_{\rm diff} = \dot{\varpi}_{\rm uqrt} - \dot{\varpi}_{\rm lqrt}$.
The choice to look at quartiles comes from our observation that if about  $\sim30\%$ of the orbits in the disk can't undergo the instability for any reason, the whole disk will fail to undergo the instability.

We define the ratio of these two timescales as
\begin{equation}
    r = \frac{t_{\rm e-fold}}{\td}.
\end{equation}
The instability should occur for $r \ll 1$ and should be suppressed for $r\gg1$. 
\deleted{Hence, $\Jc$ should correspond to $r \approx 1$.}
With numerous runs, we locate the \replaced{critical $J_2$}{$\Jc$} in simulations with different $N$, $\Md$, and orbital configurations and calculate $r(\Jc) = \rc$.

We find that $\rc$ is roughly constant with $N$ provided that $N > 100$. 
For low $N$, self-stirring within the disk causes a wide spread in semi-major axes and eccentricity which artificially amplifies differential precession. 
\deleted{We find that $\rc$ at $N=400$ and $N=800$ are consistent with each other within their errors.
Thus we expect $\lim_{N\rightarrow\infty} \rc(N)$ to be consistent with $\rc(400)$.}
In addition, we find that $\rc$ is constant with mass of the disk.
Therefore, the $\rc$ measured at $N=400, \Md = 10^{-3} M$ in the scattered disk simulations should be consistent with $\rc$ as $N\rightarrow\infty$ for all $\Md$.

We find that $\rc$ does change with the orbital configuration.
This is shown in Figure~\ref{fig:r-from-sims}. 
Here `compact' refers to the compact configuration while `sd100' and `sd250' refers to the scattered disk orbital configurations discussed in section~\ref{sec:scat-disk}.
For each configuration, $N=400$ and $\Md=10^{-3}\,M$.
From this figure, we see that $\rc \approx 1$ as expected. Large $\rc$ values mean that in one e-folding time the disk orbits have differentially precessed by more than a radian with respect to one another, meaning that this configuration is more resistant than expected to added $J_2$. Small values of $\rc$ mean that the system is less resistant, the disk orbits having precessed less than a radian in one e-folding time. 
Notably, the compact configuration is worse at resisting added $J_2$ ($\rc \sim 0.3$) than the scattered disk configurations ($\rc \sim 1$).

\begin{figure*}[!htb]
    \centering
    \includegraphics{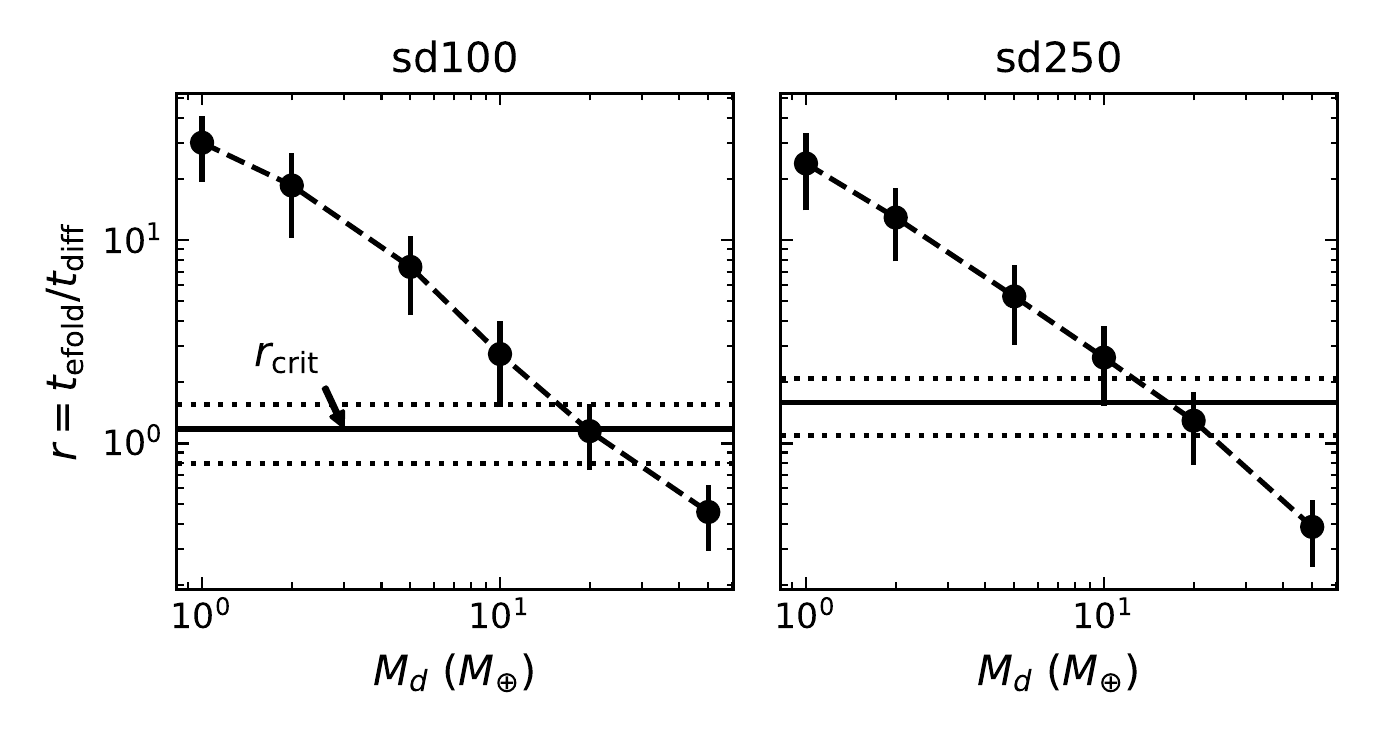}
    \caption{Ratio of e-folding to differential apsidal precession timescale, $r = t_{\rm e-fold}/\td$, in the solar system as a function of disk mass $\Md$ for two different versions of a scattered disk. The horizontal lines mark the critical timescale ratio and its error for the indicated orbital configuration as calculated from simulations (see Figure~\ref{fig:r-from-sims}). Where $r \lesssim \rc$ the instability will occur. This corresponds to $\sim20$ earth masses between 100-1000 AU.}
    \label{fig:timescale-ratio}
\end{figure*}

To scale our results to the solar system, we find the timescale ratio $r = \nicefrac{t_{\rm e-fold}}{\td}$ in the solar system for different disk masses, and compare it to the $\rc$ calculated for `sd100' and `sd250' (i.e. the points shown in Figure~\ref{fig:r-from-sims}).
The e-folding timescale for the instability in large $N$, low $\Md$, {\it compact} systems is calculated in \citet{Madigan2018}, 
	\begin{align}
	t_{\rm e-fold} \sim \chi\, \frac{0.2} {\pi} \frac{M}{\Md} P,
	\label{eq:e-fold-time}
\end{align}
where we have included a scaling factor, $\chi$, to extrapolate this result to the scattered disk configurations.
From simulations, we find that is $\chi \sim\!12$ for the 100 AU scattered disk and $\chi \sim\!11$ for the 250 AU scattered disk. That is, the e-folding timescale increases by $\mathcal{O}(10)$ accounting for the drop in mass density as particles are spread across a broad range of semi-major axes. 

We calculate the differential precession timescale directly from simulations of scattered disk configurations at the correct disk masses, $\Md = 1, 2, 5, 10, 20, 50 ~ M_{\oplus}$.
We include the $J_2$ value for the giant planets in the solar system using the appropriate semi-major axis conversion\deleted{: $a = 1 = 100~{\rm AU}$ for `sd100' ($J_2 = 5.7\times10^{-6}$), and $a = 1 = 250~{\rm AU}$ ($J_2 = 9.1\times10^{-7}$) for `sd250'}. 
Our simulation particles are fully interacting so the contribution to the precession rate from the disk potential is accounted for.
These simulations are integrated for \replaced{1000 orbits}{$\sim6\,t_{\rm sec}$ (1000 orbits)}, far too short to see the instability, but more than enough time to calculate the differential precession rate.
This rate is then used to calculate $\td$ which in combination with the $t_{\rm e-fold}$ above gives us the timescale ratio for the solar system. 

We present our results in Figure~\ref{fig:timescale-ratio}. We find that for both scattered disk configurations, $r \approx \rc$ at $\Md \approx 20 \, M_{\oplus}$.
Thus, the disk mass required for the inclination instability to occur in a primordial scattered disk in the solar system under the gravitational influence of the giant planets is $\mathcal{O}(20\,M_{\oplus})$.
For smaller disk masses, the differential precession due to giant planets suppresses the instability. 
Our previous estimate for the total mass required in the outer solar system with $a > 100$~AU for the instability to occur was about an Earth mass. \added{This estimate was based on a compact configuration, we which adopted to clearly demonstrate the discovery of a new instability in near-Keplerian disks.} 
This new estimate is $\mathcal{O}(10)$ times greater than our previous estimate demonstrating the importance of \added{both disk mass density and} differential apsidal precession in this global instability. 
Sources of error in this estimate include the unknown mass and inclination distribution as a function of radius in the primordial scattered disk. We have taken reasonable best estimates and a full exploration of parameter space is beyond the scope of the paper. 

The orbital evolution of a $\sim 20 \, M_\oplus$ mass primordial scattered disk in the solar system is modelled by our `sd100' simulations in the transition region.
Figure~\ref{fig:insta-sd-J2} shows the expected orbital evolution of a massive primordial scattered disk.
The instability saturates \replaced{by $\sim 4 \times 10^4$~P}{at $\sim250\,t_{\rm sec}$}. \replaced{Converting to solar system units using $a = 1 = 100$ AU,}{With $P=1000\,{\rm yr}$,} this is a saturation time of 40 Myr \deleted{(P = 1000 yr)}. 
\replaced{With $\Md = 6 \times 10^{-5}\,M$}{Scaling using the secular timescale}, a 20 Earth mass primordial scattered disk with the correct solar system value of $J_2$ will have a saturation timescale $\lesssim 660$~Myr.
Post-instability, the intermediate to large semi-major axis population ($a \in [200, 1000]$~AU) is extremely detached from the inner solar system with perihelia of $\approx 100 - 150$~AU. The \replaced{$a \in [200,400]$}{$a \in [200,400]\,{\rm AU}$} range actually have the largest perhelia values. 
Bodies with $a \lesssim 200$~AU will have inclinations of $i \approx30^\circ$ while bodies with $a \gtrsim 200$~AU will have inclinations twice as large, about $60^\circ$. This latter population will be very difficult to detect due to their extreme detachment. 

\section{Summary and Conclusions}
\label{sec:sum}

In this paper, we continue our exploration on the collective gravity of high eccentricity orbits in a near-Keplerian disk. We simulate the ``inclination instability'', a dynamical instability akin to buckling in barred disk galaxies which comes about from orbit-averaged torques between the individually low mass, but collectively massive, population. 
The disk orbits incline exponentially off the mid-plane, drop in eccentricity and tilt over their axes in a coherent way which leads to clustering in arguments of perihelion $\omega$. 
Starting from an unrealistic (but tractable) compact configuration of orbits, we build up to simulating a massive primordial scattered disk in the outer solar system. We include the orbit-averaged gravitational influence of the giant planets using a quadrupole $J_2$ moment of the central body. This causes the scattered disk orbits to differentially precess with respect to one another, weakening the strength of inter-orbit torques.  
We summarize our findings as follows:
\begin{enumerate}

    \item The e-folding timescale of the instability increases by $\mathcal{O}(10)$ when we simulate orbits in a scattered disk rather than in a compact configuration (Equation~\ref{eq:e-fold-time}). This is due to the drop in mass density as particles are spread across an order-of-magnitude range of semi-major axes. 

    \item We identify a critical $J_2$ moment in each simulation configuration, $\Jc$, beyond which the instability is suppressed. The growth rate of the instability decreases (by a factor of a few) with added $J_2$ moment across a transition region $(0.1 \-- 1)\,\Jc$, becomes zero at $\Jc$, and is imaginary above $\Jc$ (Figure~\ref{fig:gr-change}). 
    
    \item The median post-instability inclination/eccentricity increases/decreases with the addition of $J_2 < \Jc$ (Figure~\ref{fig:pi-orb-elem}). 
    
    \added{\item The time over which $\omega$-clustering is maintained (about 60 secular times) is not strongly affected by the addition of a $J_2$ moment for a disk in the compact configuration (Figure~\ref{fig:w-cluster}). 
    In a scattered disk configuration however, the addition of J2 increases the timescale over which $\omega$-clustering is maintained (Figure~\ref{fig:omega-cluster-sd}). The clustering persists for orbits with semi-major axes $a \gtrsim 200$ AU until the end of the simulation, about 300 secular times. }
    
    \item Physically, $J_2 = \Jc$ in a given simulation is reached when orbits in the disk precess $\approx 1$ radian apart from each other within an e-folding time (Figure~\ref{fig:r-from-sims}). Above this value, orbits differentially precess too rapidly for long-term coherent torques to sustain the instability. 
    
    \item The instability is a global phenomenon. If enough mass (and angular momentum) remains pinned to the mid-plane of the disk, the remainder of the disk is preventing from lifting off. 

    \item The mass required for the inclination instability to occur in a primordial scattered disk between $100 \sim 1000$ AU in the solar system {under the gravitational influence of the giant planets} is $\mathcal{O}(20\,M_{\oplus})$.
    We look at two different scattered disk configurations to explore the effect of distributing the peak of the mass density in a different location. Figure~\ref{fig:timescale-ratio} shows they yield the same result. 
    
    \item Unstable orbits at different semi-major axes end up with similar mean inclinations post-instability (Figures~\ref{fig:insta-sd} and \ref{fig:insta-sd-J2}). Hence, those at lower semi-major axis gain a larger fractional increase in orbital angular momentum than those at higher semi-major axis. These orbits decrease their eccentricities and increase their perihelia more so than those at higher semi-major axis. This naturally generates a perihelion gap at the innermost radius of the disk that has undergone the instability. 
    This gap appears as an under-density of orbits at perihelia of $\approx 50$ AU at semi-major axes of $\approx200 - 600$ AU. In Figure~\ref{fig:pericenter-gap}, we show the time-averaged surface density of perihelion $p$ vs semi-major axis $a$ (for all times post-instability). We need to time-average the simulation to get sufficient resolution to make this plot. This means we cannot attempt to precisely match the observed perihelion gap in the solar system \citep{Trujillo2014,Bannister2018,Kavelaars2020}. The parameters of the region depleted by the instability are related to the magnitude of $J_2$; a larger $J_2$ depletes a larger region of $a$-$p$ space.
    
    \item Orbits with semi-major axes $\approx 200 - 1000$ AU will, on average, obtain extremely large perihelion distances (p $\gtrsim100$ AU) and inclinations ($i \sim 60^\circ$). Figure~\ref{fig:pi-orb-elem} shows however that there is a broad range of final inclination and eccentricity values. It is possible to produce high inclination, high eccentricity eTNOs such as 2015 BP519  \citep{Becker2018}. 
\end{enumerate}

Now we come to the question, just how unreasonable is it to expect the primordial scattered disk to contain $\mathcal{O}(20\,M_{\oplus})$?
Current theories of planet formation suggest that the giant planets migrated significantly in a massive planetesimal disk. Scattering planetesimals fled to more stable regions of the solar system including into the various populations we observe today (hot classical resonant Kuiper belt, the scattered disk, the Trojans, irregular satellites, etc. see \citet{Nesvorny2018} for a recent review). Comets in the Oort Cloud were scattered outward, until the gravitational influence of the Galaxy could torque their orbits and detach them from the inner solar system \citep[for review see][]{Dones2015}. On their way out, they would have passed through the region of space that we are most interested in, $\approx 100 - 1000$ AU. If enough mass existed on scattered, high eccentricity orbits in this region at any given time, the orbits would collectively have gone unstable. 
    
The Nice Model of giant planet migration supposes some $30 - 50$ earth masses of planetesimals existing from the orbit of the outermost giant planet to $\sim35$ AU \citep{Gomes2005,Tsiganis2005,Morbidelli2005}.
\citet{Nesvorny2016} show that 1000-4000 Pluto mass bodies are needed in a primordial outer planetesimal disk to match the observed current population of Neptune's resonant bodies. Their estimate is obtained by modelling Neptune's migration through this disk. The authors find that a primordial disk of $\sim20$ Earth masses is consistent with the observed resonant populations \citep[see also][]{Nesvorny2012}. 

More recently, \citet{Shannon2018} estimate the mass of the primordial scattered disk using the survival of ultra-wide binaries in the Cold Classical Kuiper belt. At a 95\% upper limit they find the disk could have contained 9 Earths, 40 Mars, 280 Lunas, and 2600 Plutos (at the 68\% upper limit the numbers are 3, 16, 100, 1000). The combined mass of just these bodies is $\sim$23 Earth masses ($\sim$8 Earth masses at 68\%). The total mass of the disk would be significantly more than this.

If at any point 20 Earth masses of material scattered onto high eccentricity orbits between $\approx 100 - 1000$ AU, it would have undergone the inclination instability. 
Now we turn to the question of how much of this mass would remain today.  
The total mass remaining in this region depends critically on the outgoing flux from this population, or how many bodies have been lost over the age of the solar system. Post-instability, the population is relatively isolated as 
the orbits drop in eccentricity at the same time they incline off the ecliptic. This raises their perihelia and lowers their aphelia, reducing the influence of the inner solar system planets on one side, and galactic tides and passing stars on the other. The population fossilizes at high inclinations and  extraordinary values of perihelion distances. 
The outgoing flux should therefore depend on secular gravitational interactions between the orbits themselves rather than outside influences. These secular torques in turn depend on the distribution of the eTNOs today, particularly on whether or not they align in physical space \citep{Sefilian2019,Zderic2020}. 
The secular gravitational torques will cause long-term angular momentum changes transferring some bodies from the large detached population into and around the current-day scattered disk.

\citet{Hills81} provides an early estimate for the current amount of mass in the region spanning the orbit of Neptune to $10^4$~AU. Using a variety of heuristic arguments involving the Oort cloud, Hills suggests there could be anywhere from a few to a few thousand Earth masses of material (an average yields tens of Earth masses). 

\citet{Hogg1991} derives a current estimate by considering the dynamical influence of a massive ecliptic disk with $a \gg 30$~AU on the ephemerides of the giant planets and Halley's comet. They find that there could be hundreds of Earth masses of material in the disk based on the giant planets ephemerides or a few Earth masses of material based on the ephemerides of Halley's comet. 
Hogg's latter limit is the more constraining estimate, suggesting that the primordial scattered disk population, if it exists, must be whittled down to a few Earth masses. It's worth noting however that they assume a flat disk with the potential modelled as an outer quadrupole moment. 

\citet{Gladman2009} report the discovery of a TNO on a retrograde orbit and find the object is unlikely to be primordial. They suggest a supply mechanism from a long-lived source, for example a population of large-inclination orbits beyond Neptune. This could also be a source for Halley-type comets \citep{Levison2006}. The inclination instability in a primordial scattered disk produces such a reservoir.
As posited in \citet{Madigan2016}, there may be a massive ($1$ - $10\,M_{\oplus}$) reservoir of icy bodies at large orbital inclinations beyond the Kuiper Belt. The Vera C. Rubin Observatory \citep{Ivezic2019} will be instrumental in the discovery and orbital classification of this population should it exist. 
    
Finally, the results we present here focus on the instability during or soon after its linear phase. 
However, a 20 Earth mass primordial scattered disk has a saturation timescale $\lesssim 660$~Myr. If the outer solar system did undergo this instability, it must be in the non-linear, saturated state. The long-term behavior of the instability is therefore crucial to understand in order to compare with current-day observations.

\acknowledgments
This work was supported from NASA Solar System Workings under grant 80NSSC17K0720 and NASA Earth and Space Science Fellowship 80NSSC18K1264. AM gratefully acknowledges support from the David and Lucile Packard Foundation.
This work utilized the RMACC Summit supercomputer, which is supported by the National Science Foundation (awards ACI-1532235 and ACI-1532236), the University of Colorado Boulder, and Colorado State University. The Summit supercomputer is a joint effort of the University of Colorado Boulder and Colorado State University.

\software{\texttt{REBOUND}  (Rein and Liu 2012, Rein and Spiegel 2015)}

\appendix

\deleted{
\section{Precession due to quadrupole potential}
\label{app:J2}
Our aim is to find the secular precession rate of an orbit due to a quadruple potential. 
Starting with Newton's force law,
\begin{equation}
\ddot{\bm{r}} = -\nabla\Phi,
\end{equation}
we consider $\Phi = \Phi_{\rm Kep} + \mathcal{R}$ where $\mathcal{R}$ is some perturbation on the otherwise Keplerian potential. In orbital perturbation theory, $\mathcal{R}$ is called the disturbing function. This becomes,
\begin{equation}
\ddot{\bm{r}} + \mu \frac{\bm{r}}{r^3} = \nabla \mathcal{R},
\end{equation}
where $\mu = GM$.  For our two term multipole expansion (equation \ref{eq:multipole}), $\mathcal{R}$ equals the quadrupole term in the potential,
\begin{equation}
\mathcal{R} = -J_2 \frac{\mu R^2}{r^3} \, P_2\left(\cos{\theta}\right). 
\end{equation}
Lagrange's planetary equations give us the time evolution of the osculating Keplerian orbital elements in terms of partial derivatives of the disturbing function. The equations for $\dot{\omega}$ and $\dot{\Omega}$ are,
\begin{align}
    \frac{d\omega}{dt} &= \frac{\sqrt{1-e^2}}{na^2e}\,\frac{\partial \mathcal{R}}{\partial e} - \frac{\cot{i}}{na^2\sqrt{1-e^2}}\,\frac{\partial \mathcal{R}}{\partial i}, \\
\frac{d\Omega}{dt} &= \frac{1}{na^2\sin{i}\sqrt{1-e^2}}\,\frac{\partial \mathcal{R}}{\partial i}. 
\end{align}
Our next steps are to find $\mathcal{R}$ in terms of the Keplerian orbital elements, average $\mathcal{R}$ over an orbit and plug it into the Lagrange planetary equations for $\dot{\omega}$ and $\dot{\Omega}$ to find the secular precession rate.

We begin by noting that,
\begin{equation}
\cos{\theta} = \frac{z}{r} = \sin{\left(\omega + f\right)}\sin{i},
\end{equation}
where $z$ is the standard Cartesian coordinate and $f$ is the true anomaly. Using the expression for the Legendre polynomial, $P_2(x) = \nicefrac{1}{2}(3x^2-1)$, and a trig. identity, $\mathcal{R}$ becomes,
\begin{equation}
\mathcal{R} = \frac{J_2}{2} \, \frac{\mu R^2}{r^3} \left[ 1 - \frac{3}{2}\sin^2i + \frac{3}{2}\sin^2i \cos{\left( 2f + 2\omega \right)} \right].
\end{equation}
We know $r$ as,
\begin{equation}
r = \frac{a\left(1-e^2\right)}{1+e\cos f},
\end{equation}
which makes $\mathcal{R}$
\begin{equation}
\mathcal{R} = \frac{J_2}{2} \, \frac{\mu R^2 \left(1+e\cos f\right)^3}{a^3\left(1-e^2\right)^3} \left[ 1 - \frac{3}{2}\sin^2i + \frac{3}{2}\sin^2i \cos{\left( 2f + 2\omega \right)} \right] .
\end{equation}
Now that we have $\mathcal{R}$ in terms of the Keplerian elements we average over the orbital period,
\begin{equation}
 \overline{\mathcal{R}} = \frac{1}{P} \int_0^P \mathcal{R} dt. 
 \end{equation}
This can be converted to an integral over $f$ by considering that equal areas are swept out in equal time in a Keplerian orbit,
\begin{align}
{\rm const} &= \frac{dA}{dt} = \frac{1}{2}r^2\frac{df}{dt} = \frac{\pi a^2 \sqrt{1-e^2}}{P}, \\
dt &= \frac{r^2}{a^2} n^{-1} \left(1-e^2\right)^{-\nicefrac{1}{2}} df. 
\end{align}
Making this change of variables and using the expression for $r$,
\begin{equation}
\overline{\mathcal{R}} = \frac{\left(1-e^2\right)^{\nicefrac{3}{2}}}{2 \pi} \int_0^{2\pi} \frac{\mathcal{R}}{\left(1+e\cos{f}\right)^2} \, df. 
\end{equation}
Next, we insert our expression for $\mathcal{R}$,
\begin{equation}
\overline{\mathcal{R}} = \frac{J_2}{4\pi} \, \frac{\mu R^2}{a^3\left(1-e^2 \right)^{\nicefrac{3}{2}}} \int_0^{2\pi} \left(1+e\cos{f}\right)\left( 1 - \frac{3}{2}\sin^2i + \frac{3}{2}\sin^2i \cos{\left( 2f + 2\omega \right)} \right) df. 
\end{equation}
Though busy, this integral is straight-forward,
\begin{equation}
\overline{\mathcal{R}} = \frac{J_2}{2} \, \frac{\mu R^2}{a^3\left(1-e^2 \right)^{\nicefrac{3}{2}}} \left(1 - \frac{3}{2}\sin^2{i} \right).
\end{equation}
Now that we have the orbit-averaged disturbing function, can use the Lagrange planetary equations to find the secular evolution of $\omega$ and $\Omega$. Plugging in $\overline{\mathcal{R}}$ and simplifying, we get,
\begin{equation}
\dot{\omega} = \frac{3J_2}{4} \, n \, \frac{R^2}{a^2} \frac{5\cos^2{i} - 1}{\left(1-e^2\right)^2}, \qquad \text{and,} \qquad \dot{\Omega} = \frac{3J_2}{2} \, n \, \frac{R^2}{a^2} \, \frac{\cos{i}}{\left(1-e^2\right)^2}, 
\end{equation}
which can be combined to give $\dot{\varpi}$.}

\section{Precession due to the disk mass}
\label{app:disk}

\begin{figure*}
    \centering
    \includegraphics{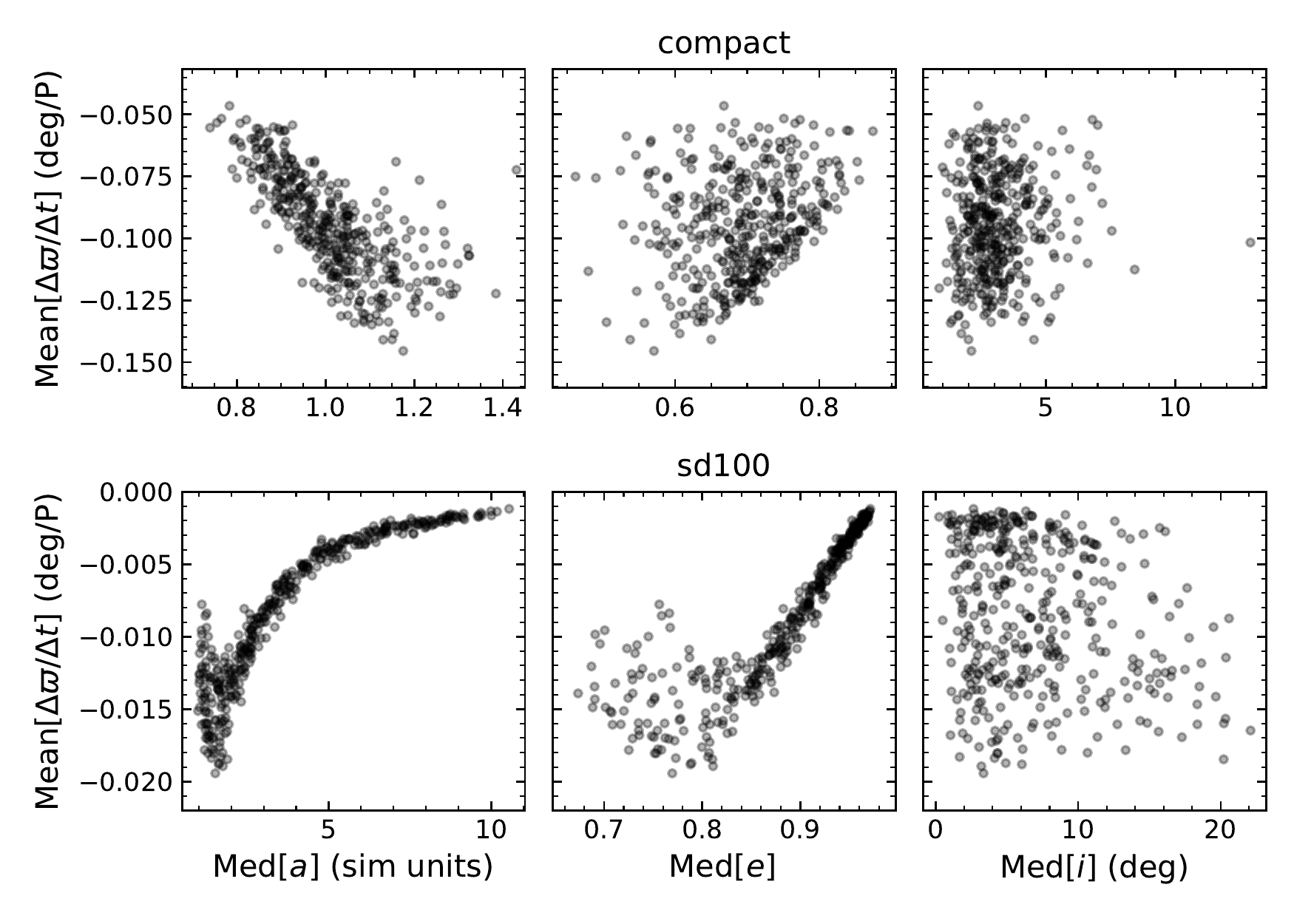}
    \caption{Median apsidal precession rate for bodies in a compact configuration (top row) and scattered disk (bottom row) simulation. The apsidal precession rate is always retrograde in the disk and is not a strong function of inclination $i$. (Top Row) Semi-major axis $a$ is the primary source of differential precession. Although the $a$ dependence seems linear, the range is so narrow that the plot reflects a Taylor expansion of the true dependence at $a=1$. Note that the apsidal precession dependence on $a$ changes for $a \gtrsim 1.1$ from linearly decreasing to increasing. 
    (Bottom Row) The apsidal precession rate is again retrograde and a strong function of semi-major axis. There are two distinct regimes. With $a$ in $[1,2]$, the retrograde apsidal precession rate increases in magnitude with $a$ then in $[2,10]$ decreases towards zero. 
    }
    \label{fig:prec-vs-orb-elem}
\end{figure*}

\begin{figure*}[!htb]
    \centering
    \includegraphics{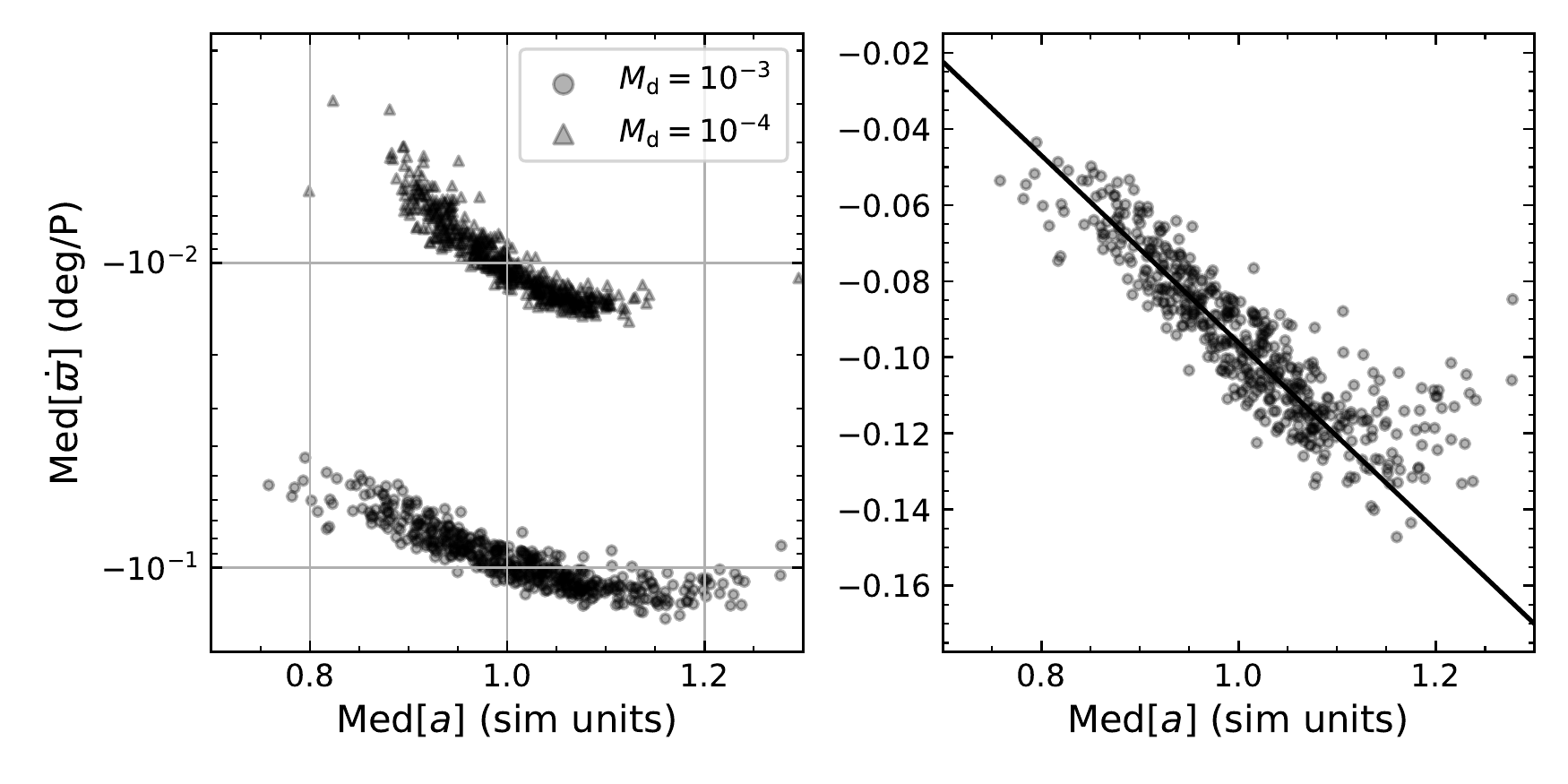}
    \caption{Apsidal precession rate in the compact orbital configuration (pre-instability) with $N=500$, as a function of  semi-major axis. (Left) We show the precession rate for two different initial disk masses, $\Md$. The magnitude of the precession rate scales $\sim$linearly with the mass of the disk and the precession rate is retrograde. (Right) Apsidal precession rate for the case where $\Md = 10^{-3}\,M$ with a linear $y$ scale and a simple fit. The precession rate varies $\sim$linearly with $a$, with the outer edge of the disk precessing faster than the inner edge. On the edges of the disk, the apsidal precession rate doesn't quite follow this linear dependence. Note, only the particles with average semi-major axis in the range [0.9, 1.1] were factored into the shown fit. This linear functional form is conserved over changes in $\Md$, and the slope of the fit increases as disk mass is decreased.}
    \label{fig:compact-prec}
\end{figure*}

\begin{figure*}[!htb]
    \centering
    \includegraphics{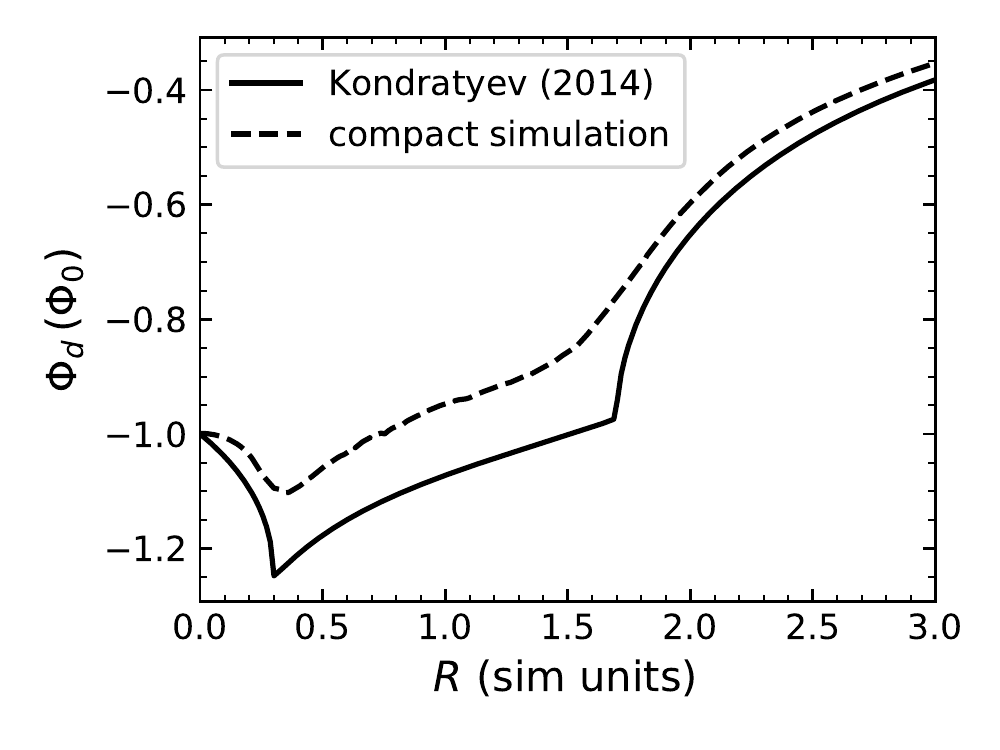}
    \caption{Disk potential in the $xy$-plane as a function of cylindrical radius $R$ normalized to the value of the disk potential at $R=0$. The potential for a disk of eccentric orbits derived in \citet{Kondratyev2014} is shown with $a=1$ and $e=0.7$ along with the potential of a simulated disk initialized in the compact configuration at $t=100\,{\rm P}$. The disk potential calculated from the simulation is `softened' in the sense that each orbit in the disk has been sampled at 40 evenly-spaced mean anomaly points. The general form of the potential is the same for both with the simulation potential lacking the cusp at $R=0.3$ and $R=1.7$ of the Kondratyev potential due to the small spread in $a$ and $e$ that the disk has naturally developed from two-body scattering.}
    \label{fig:disk-pot}
\end{figure*}

Here we describe apsidal precession of orbits due to the potentials of the pre-instability (relatively flat) disks presented in this paper. 
First, we look at numerical results from simulations showing how the rate of change of longitude of perihelion, $\dot{\varpi}$, varies with semi-major axis, eccentricity, and inclination for the compact and scattered disk (sd100) configurations (Figure~\ref{fig:prec-vs-orb-elem}), and \deleted{how it scales with} mass of the disk (Figure~\ref{fig:compact-prec}).
Second, we compare the potential of the disk in the compact configuration to an analytic expression derived in \citet{Kondratyev2014}. 
This is shown in Figure~\ref{fig:disk-pot}.
We discuss how this potential can be used to explain some features of the compact and scattered disk apsidal precession profiles.

In Figure~\ref{fig:prec-vs-orb-elem}, we show median apsidal precession rate vs.\ median semi-major axis, eccentricity, and inclination  for both the compact and `sd100' orbital configurations with the same disk mass, $\Md = 10^{-3} M$, and number of particles $N = 400$.
The magnitude of the median precession rate in the compact configuration is approximately 10 times higher than the median precession rate in the scattered disk configuration, reflecting the lower mass density in the latter. 
In both cases, apsidal precession is retrograde and inclination $i$ has minimal effect on $\dot{\varpi}_d$.
In the compact configuration, the magnitude of the precession rate increases with semi-major axis. 
This relationship is roughly linear.
In the scattered disk configuration, the magnitude of the precession rate increases from $a\sim1$ to $a\sim2$ after which it decreases out to $a\sim10$.
The sum of $\dot{\varpi}_d$ for the scattered disk and $\dot{\varpi}_{J2}$ results in a flattened precession profile (at least for $a > 2$), while the sum of $\dot{\varpi}_d$ for the compact configuration and $\dot{\varpi}_{J2}$ results in a steeper profile.

Figure~\ref{fig:compact-prec} shows the disk-only precession rate of orbits in the compact configuration pre-instability with slightly higher $N$ (500 vs.\ 400 for Figure~\ref{fig:prec-vs-orb-elem}). 
In the left panel, we see the secular scaling of the precession rate ($\dot{\varpi}_d \propto \Md$).
In the right panel, we see the linear dependence of $\dot{\varpi}$ on $a$. 
This dependence is clearer here due to the increased $N$.

In the compact orbital configuration, the potential of the disk is well approximated by an expression derived in \citet{Kondratyev2014}.
Kondratyev found the potential of an infinitely-populated axisymmetric disk of orbits with zero inclination and equal semi-major axis and eccentricity, $a_d$ and $e_d$. 
This solid washer mass distribution is characterized by the inner and outer radii, $R_1 = a_d (1-e_d)$ and $R_2 = a_d(1+e_d)$. 
The radial mass density of the washer is the inverse of the radial Kepler velocity.
The resulting potential in the plane of the washer is piece-wise, and expressed as integrals over the mass distribution,
\begin{equation}
    \label{eq:kond-pot}
    \Phi_d(R) = \frac{2 \phi_0}{\pi^2}
    \begin{dcases*} 
        \int_{R_1}^{R_2} \sigma(x) K\left( \frac{R}{x} \right) dx & \text{if $R <     R_1$} \\
        \frac{1}{R} \int_{R_1}^{R} x \sigma(x) K\left( \frac{x}{R} \right) dx +     \int_{R}^{R_2} \sigma(x) K\left( \frac{R}{x} \right) dx & \text{if      $R_1 < R < R_2$} \\
        \frac{1}{R} \int_{R_1}^{R_2} x \sigma(x) K\left( \frac{x}{R} \right) dx &     \text{if $R > R_2$}
    \end{dcases*} 
\end{equation}
where $K$ is the complete elliptic integral of the first kind, $\phi_0 = -\nicefrac{G\Md}{a_d}$ is the potential at the origin, and $\sigma(x) = ((R_2 - x)(x - R_1))^{-\nicefrac{1}{2}}$. 

This potential is shown in Figure~\ref{fig:disk-pot} along with the potential of a compact configuration disk with $N=400$ at $t=100~{\rm P}$. 
Each orbit was sampled at 40 equally-spaced mean anomalies and the potential was averaged along 10 different azimuthal lines in the $xy$-plane. 
Note that we expect the simulation potential to differ from the Kondratyev expression because of the (small) initial spread in $a$, $e$, and $i$. Despite the differences, the two potentials share the same bulk characteristics. 

A formal expression for the precession rate of the orbits in the disk can be found using a Hamiltonian approach.
Restricting ourselves to the $xy$-plane, the modified Delaunay coordinates in 2D are \citep{MorbidelliMCM},
\begin{align}
    & \lambda = \mathcal{M} + \varpi & I&=\sqrt{\mu a}, \\
    &\varpi & K&=\sqrt{\mu a}( \sqrt{1-e^2} - 1).
\end{align}
We can then use Hamilton's equation's to get the time evolution of the apsidal angle (see \citet{MerrittGalacticDynamics} for a similar derivation),
\begin{equation}
    \dot{\varpi} = \frac{\partial H}{\partial K},
\end{equation}
where $H$ is the Hamiltonian of the system,
\begin{equation}
    H = H_{\rm kep} + H_{d},
\end{equation}
where $H_{\rm kep} \gg H_{d}$.
The Keplerian Hamiltonian is
\begin{equation}
    H_{\rm kep} = - \frac{1}{2}\left( \frac{GM}{I} \right)^2,
\end{equation}
and we average the disk potential over a Keplerian orbit, such that, 
\begin{equation} 
    H_{d} = \overline{\Phi}_{d},
\end{equation}
where the over-line denotes an average over the unperturbed orbit and $\Phi_{d}$ is the potential of disk.
Thus, the apsidal precession rate in the disk is given by,
\begin{equation}
    \dot{\varpi}_{d} = \frac{\partial \overline{\Phi}_{d}}{ \partial K} = - \sqrt{\frac{1-e^2}{\mu a e^2}} \, \frac{\partial \overline{\Phi}_{d}}{ \partial e},
\end{equation}
where $\overline{\Phi}_{d}$ is given by the average of Kondratyev's potential, equation~\ref{eq:kond-pot}, over the unperturbed orbit. 
The average over the orbit is
\begin{equation} 
    \overline{\Phi}_{d} = \frac{1}{2\pi} \int_{0}^{2\pi} dE \, (1-e\cos{E}) \, \Phi_{d}(r),
\end{equation}
where $E$ is the eccentric anomaly. $r$ and $E$ are related by $r(E) = a(1-e\cos{E})$. We can pull the partial derivative inside the integral, use integration by parts, and the $r(E)$ expression to obtain,
\begin{equation}
    \frac{\partial \overline{\Phi}_{d}}{\partial e} = \frac{1}{\pi} \, \int_{a(1-e)}^{a(1+e)} dr \frac{e^2 - (1-r/a)}{\sqrt{e^2 - (1-r/a)^2}} \frac{d\Phi_d}{dr},
\end{equation}
\begin{equation}
    \label{eq:varpi-dot}
    \dot{\varpi}_{d} = -\frac{1}{\pi} \, \sqrt{\frac{1-e^2}{\mu a e^4}} \, \int_{a(1-e)}^{a(1+e)} dr \frac{e^2 - (1-r/a)}{\sqrt{e^2 - (1-r/a)^2}} \frac{d\Phi_d}{dr}.
\end{equation}
There is no convenient expression for the derivative of Kondratyev's potential, so we will not attempt to find a closed form expression for the apsidal precession rate.
However, we can use equation~\ref{eq:varpi-dot} along with Figure~\ref{fig:disk-pot} to understand some basic features of the precession rate shown in figure~\ref{fig:compact-prec}.
Kondratyev's potential has cusps at $R_1$ and $R_2$.
Further, Kondratyev's potential is concave down for all $R$ ($\nicefrac{d^2\Phi_d}{dR^2} < 0$).
The true disk potential does not have these cusps because the disk orbits have a range in $a$, $e$, and $i$.
Instead the disk potential has a region where the potential is concave up ($\nicefrac{d^2\Phi_d}{dR^2} > 0$) near $R_1$ and $R_2$.
In these regions $\nicefrac{d\Phi_d}{dR}$ increases with $R$.
The disk orbits are sufficiently eccentric that the orbit-averaged slope of the potential (i.e. the integral in equation~\ref{eq:varpi-dot}) is approximately given by the value of $\nicefrac{d\Phi_d}{dR}$ at apocenter. 
The orbits with $a \lesssim 1.1$ have apocenters near $R_2$ in the region where the potential is concave up.
The slope of the potential here is positive, and increasing with $a$ (assuming $e \approx 0.7$).
Thus, we would expect the apsidal precession rate of the orbits in the disk to be retrograde with magnitude increasing with $a$ until $a \approx 1.1$.
This is precisely what we see in Figure~\ref{fig:compact-prec}.

\bibliography{main.bib}

\end{document}